\documentclass[lineno]{jfm}
\usepackage{graphicx}
\usepackage{newtxtext}
\usepackage{newtxmath}
\usepackage{natbib}
\usepackage{xcolor}
\usepackage{fdsymbol}
\usepackage{pifont}
\usepackage{hyperref}
\usepackage{url}
\usepackage{float}
\usepackage{soul}
\usepackage{tikz}

\newcommand{\RomanNumeralCaps}[1]
\linenumbers
\hypersetup{colorlinks = true, urlcolor = blue, citecolor = black}
\definecolor{k1}{rgb}{0.75, 0.75, 0.75}
\definecolor{k2}{rgb}{0.60, 0.60, 0.60}
\definecolor{k3}{rgb}{0.45, 0.45, 0.45}
\definecolor{k4}{rgb}{0.30, 0.30, 0.30}
\definecolor{k5}{rgb}{0.00, 0.00, 0.00}

\definecolor{r1}{rgb}{1.00 ,0.75, 0.75}
\definecolor{r2}{rgb}{1.00, 0.60, 0.60}
\definecolor{r3}{rgb}{1.00, 0.45, 0.45}
\definecolor{r4}{rgb}{1.00, 0.30, 0.30}
\definecolor{r5}{rgb}{1.00, 0.00, 0.00}

\definecolor{b1}{rgb}{0.75 ,0.75, 1.00}
\definecolor{b2}{rgb}{0.60, 0.60, 1.00}
\definecolor{b3}{rgb}{0.45, 0.45, 1.00}
\definecolor{b4}{rgb}{0.30, 0.30, 1.00}
\definecolor{b5}{rgb}{0.00, 0.00, 1.00}

\definecolor{g1}{rgb}{0.75 ,0.90, 0.75}
\definecolor{g2}{rgb}{0.60, 0.84, 0.60}
\definecolor{g3}{rgb}{0.45, 0.78, 0.45}
\definecolor{g4}{rgb}{0.30, 0.72, 0.30}
\definecolor{g5}{rgb}{0.00, 0.60, 0.00}

\newcommand{\zp}{z^{+}}
\newcommand{\up}{U^{+}}
\newcommand{\ut}{U_{\tau}}
\newcommand{\ret}{Re_{\tau}}
\newcommand{\ie}{i.e.\ }
\newcommand{\egg}{e.g.\ }

\definecolor{OliveGreen}{RGB}{0,200,0}
\definecolor{DarkBlue}{RGB}{0,0,200}

\newcommand{\az}[1]{{\color{DarkBlue}#1}}

\title{High-Reynolds-number turbulent boundary layers under adverse pressure gradients. Part 1. Decoupling local and upstream pressure gradient effects}
\author{Ahmad Zarei, Mitchell Lozier, Rahul Deshpande \and Ivan Marusic \corresp{\email{imarusic@unimelb.edu.au}}}
\affiliation{Department of Mechanical Engineering, The University of Melbourne, Victoria 3010, Australia}

\begin{document}
\maketitle

\begin{abstract}
This study reports a carefully controlled examination of the universality of the von Kármán and additive coefficients associated with the classical logarithmic scaling law for the mean streamwise velocity profile in high-friction-Reynolds-number ($\ret$) turbulent boundary layers (TBLs) subjected to low-to-moderate adverse pressure gradients (APGs). 
The present approach leverages a recently developed methodology for selectively prescribing pressure gradients (PGs) along the full test section of Melbourne’s high-$\ret$ boundary layer wind tunnel (\citeauthor{deshpande}, \textit{Phys. Rev. Fluids}, vol.\ 8, 2023), in combination with direct measurements of the local friction velocity via oil-film interferometry. 
This methodology enables systematic variation of the upstream PG history, e.g. introducing weak upstream APG perturbations or maintaining nominally zero-pressure-gradient (ZPG) conditions (a `minimal' PG history), while obtaining locally matched $\ret$ and $\beta$ at the downstream measurement locations. 
As such, this unique experimental configuration allows for systematic decoupling of the influences of $\ret$, local APGs, and PG history, and assessment of their individual contributions to single-point turbulence statistics and energy spectra across different regions of the TBL. 
Due to the high-$\ret$ and low-to-moderate APG conditions utilized in this study, the overlap region of the TBL was well-defined and had a sufficient extent in the mean streamwise velocity profile to rigorously examine the classical logarithmic scaling law. 
The von Kármán coefficient was found to remain invariant across all cases, within experimental uncertainty, while the additive coefficient varied systematically with both the local APG and PG history, possibly explaining variabilities in the logarithmic scaling behaviours reported across the APG TBL literature. 
In addition to the clear separation between the inner and wake regions, the present high-$\ret$ experiments provide sufficient scale separation to resolve large- and small-scale contributions and distinguish their responses to local APG and PG history.
Local APGs were observed to energize both large- and small-scale motions in the wake region around $0.4\delta$, as expected, while PG history was found to also influence large-scale motions extending to approximately $0.25\delta$, just above the overlap region. 
In contrast to previous lower-$\ret$ studies, neither effect was observed to extend into the inner region, highlighting the importance of the scale/region separation achieved here. 
Collectively, these measurements provide a high-fidelity dataset that reveals the underlying flow physics, improving fundamental understanding and providing the basis for developing composite mean velocity profile formulations for APG TBLs that account for the distinct influences of $\ret$, local PGs, and PG history.
\end{abstract}

\section{Introduction}
\label{Intro}

Turbulent boundary layers (TBLs) subjected to adverse pressure gradients (APGs) are relevant to a variety of engineering applications, such as the flows over airfoils or within diffusers. 
The relevant properties of these APG TBLs are governed by specific flow conditions, namely the local friction Reynolds number, the local streamwise pressure gradient (PG) and the \emph{history} of streamwise PGs experienced by the TBL upstream of the measurement location \citep{devenport2022equilibrium}. 
The local friction Reynolds number, $\ret=\delta\ut/\nu$, is a fundamental parameter of the TBL which describes (exclusively) the properties of well-behaved, smooth-wall, zero-pressure-gradient (ZPG) TBLs \citep[\ie canonical TBLs;][]{marusic_evolution_2015, vila2017identification}. 
Here, $\delta$ denotes the boundary-layer thickness \citep{lozier2025defining}, $\ut$ is the mean friction velocity, and $\nu$ is the kinematic viscosity. 
Local streamwise PGs will be quantified in this study using Clauser’s pressure-gradient parameter, $\beta=(\delta^{*}/\rho U_{\tau}^{2})(dP/dx)$, where $\delta^{*}$ is the displacement thickness, $\rho$ is the density, and $dP/dx$ is the local streamwise pressure gradient, and $x$ and $z$ denote the streamwise and wall-normal directions, respectively. 
Invariably, however, local PGs are generally realized by varying the upstream flow conditions, leading to coupled PG history effects.
While the influences of the local flow conditions (\ie $\ret$ and $\beta$) on APG TBL properties, like the mean turbulence statistics, have been studied over many decades \citep{clauser_turbulent_1954}, fewer studies have sought to systematically \emph{decouple} and quantify the additional influence of PG history. 
For instance, it is well established that local streamwise PGs influence the magnitude of turbulent stresses (\egg the emergence of secondary peaks in the wake region; \citealp{skaare1994turbulent, marusic1995wall, deshpande2024}) and the mean streamwise velocity \citep[\egg increase in the wake factor, $\Pi$, with $\beta$;][]{monty2011}. 
However, a number of recent studies \citep{devenport2022equilibrium} have demonstrated that PG histories can influence these same turbulence statistics, on par with the influence of local PGs, motivating the need for further studies to decouple these effects. 

In particular, the coupled influences of these flow parameters have historically made it difficult to rigorously characterize, solely, the influence of local APGs on the classical scaling laws for TBL statistics \citep{monty2011,pozuelo,nagib_variations_2008}. 
An example is that of the classical logarithmic-law, which describes the mean velocity profile in the overlap region of the TBL, and is given by
\begin{equation}
    \up = \frac{1}{\kappa} \ln (\zp) + B,
    \label{eq:loglaw}
\end{equation}
where $\kappa$ and $B$ are the von K\'arm\'an and the additive coefficients, respectively. 
Despite the widespread acceptance of this logarithmic scaling behaviour in the overlap region of high-$\ret$ canonical TBLs \citep{marusic_logarithmic_2013, smits2011high}, there remain unresolved questions about if, and how, $\kappa$ and $B$ are influenced by local APGs and PG history individually \citep{nickels, nagib_variations_2008, knopp2021experimental}. 
\az{Consequently, the aim of the present study is to systematically decouple the influences of local APGs and PG history to determine unambiguously how local APGs influence these classical log-law coefficients.} 
In this study, $\kappa = 0.39$ and $B = 4.3$ will be used for canonical TBLs \citep{marusic_evolution_2015}, and the “+” superscript will be used to denote normalization by viscous velocity ($\ut$), length ($\nu/\ut$), and/or time ($\nu/\ut^2$) scales. 

\begin{figure}
  \captionsetup{width=1.00\linewidth}
   \begin{center}
    \includegraphics[width=1.00\textwidth]{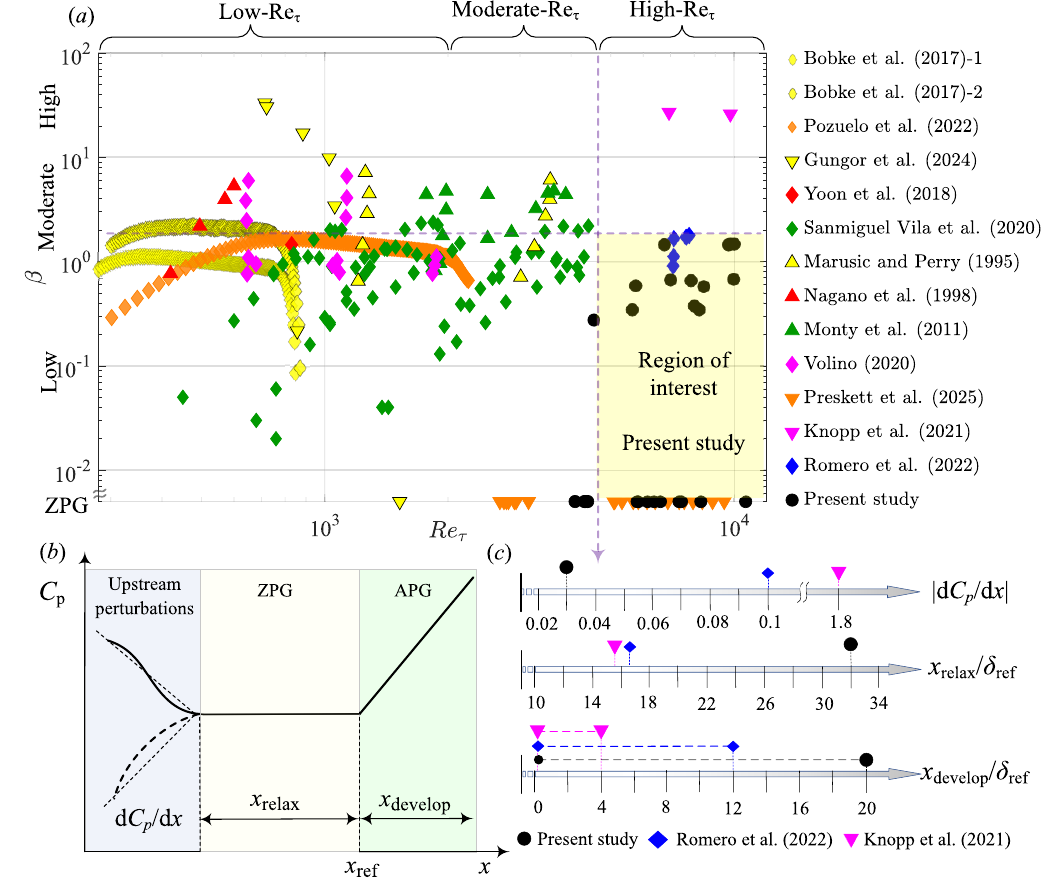}
   \end{center}
   \caption{\az{(\textit{a}) $\beta(\ret)$ map of APG TBLs datasets from the literature, including numerical and experimental studies. The yellow-shaded area represents the region of interest for the present study, i.e. high $\ret$ and low-to-moderate $\beta$ where a well-defined overlap region is expected. (\textit{b}) Schematic representation of the different regions of TBL development typically encountered in high-$\ret$ APG experiments. (\textit{c}) Comparison of experimental parameters between the present study and those reported by \citet{romero_properties_2022} and \citet{knopp2021experimental}. $\delta_{\rm{ref}}=\delta(x=x_{\rm{ref}})$.}}
   \label{fig1}
\end{figure}

\subsection{\az{Investigation of the classical logarithmic-law coefficients}}

There are a number of considerations, outlined here, that help ensure investigations of the classical log-law in APG TBLs are rigorous and systematic.
First, the classical scaling laws for mean TBL statistics are observed at sufficiently high Reynolds numbers, where a clear separation exists between the inner and wake regions of the TBL, and where the extent of the overlap region permits unambiguous evaluation of these laws \citep{smits2011high}. Here, we consider the higher $\ret$ regime to correspond to $\ret \gtrsim 4500$, consistent with \citet{marusic_logarithmic_2013}, as shown in figure~\ref{fig1}(\textit{a}).
Along similar lines, strong local pressure gradients (defined here as $\beta \gtrsim O(10)$; see figure~\ref{fig1}\textit{a}) have also been found to influence the extent of the overlap region. Specifically, increasing local APG strength corresponds to a \emph{decrease} in the upper limit of the overlap region \citep{marusic1995wall, knopp2021experimental}. 
While many studies have examined low-$\ret$ TBLs and/or TBLs subjected to strong APGs (e.g. leading to separation), which are relevant to many applications, such conditions are not well suited for evaluating the log-law. 
The combination of limited scale separation and/or strong local APG effects can severely compress (or obscure) the overlap region, thereby introducing uncertainty in reported variations of $\kappa$ and/or $B$. 
Another consideration, particularly for PG TBLs, is the influence of PG history. For example, changes in $B$ have previously been linked to both $\beta$ \citep{nagano1998structure, monty2011} and PG history effects \citep{bobke2017history, gungor2024turbulent, preskett2025effects}.
Because real-life applications almost always involve sources of PG history (\eg\ airfoil geometries), many studies have been designed to introduce targeted PG-history effects at both low and high $\ret$ \citep{bobke2017history, gungor2024turbulent, preskett2025effects}. 
In such cases, $\kappa$ and $B$ may be influenced by both $\beta$ and the imposed PG histories, which are unique to each study. 
Decoupling these influences, in order to isolate the effects of either $\beta$ or PG history alone, requires additional cases and measurements with systematic, well-controlled variations in PG history and local flow conditions.

\az{To highlight the significance of the experimental dataset evaluated in the present study, we compare the corresponding parametric space with that of previous investigations. The $\beta(\ret)$ maps for a selection of previous APG TBL datasets are presented in figure~\ref{fig1}(\textit{a}) for reference. 
This figure illustrates the wide range of $\ret$, $\beta$, and PG-history conditions that have already been investigated, both numerically and experimentally.} 
However, in the parameter space best suited to investigate the log-law coefficients (\ie the region of high $\ret$ and low-to-moderate $\beta$ highlighted in yellow in figure~\ref{fig1}a) only one experimental study has been conducted to investigate the scaling behaviours of APG TBL statistics (at least in part). 
Importantly, the APG conditions in all past experiments have also been generated using different methods, namely adjustable converging and/or diverging ceilings \citep{volino_non-equilibrium_2020, parthasarathy2023family, sanmiguel_vila_experimental_2020}, ramp setup \citep{romero_properties_2022, knopp2021experimental}, or wall-normal transpiration \citep{clauser_turbulent_1954, deshpande}. 
A generalized schematic of the streamwise PG variations, which are representative of these past APG TBL experiments, is shown in figure~\ref{fig1}(b). 
In general, the most upstream region often experiences favourable-pressure-gradient (FPG) conditions, introduced by the geometry of ramp or ceiling \citep[see for instance][]{sanmiguel_vila_experimental_2020,romero_properties_2022,knopp2021experimental}. 
Although these upstream PG conditions are highly setup/facility-dependent, any non-zero PGs in the farthest upstream region are referred to and interpreted here as `perturbations', because they can potentially influence the downstream development of the TBL. 
These upstream PG perturbations are typically followed by a nominally ZPG `relaxation' region, designed to damp the effects of these perturbations and re-establish a canonical TBL state. 
APG conditions are then introduced farther downstream in a `development' region where the APG TBLs are typically measured. 
While most experimental designs include a relatively long relaxation region, to minimize upstream PG history effects and ensure a well-defined baseline prior to the onset of APG conditions, each experimental configuration is associated with a \emph{distinct} PG history. 
This variability/uniqueness of each setup complicates direct comparisons between different APG TBL datasets, as it is uncertain if the turbulence statistics in the development region retain any influences from the specific PG histories imposed upstream. This underscores the importance of investigating APG TBL with minimal upstream history, which can be interpreted as a `canonical' condition for understanding PG TBL scaling.

Additionally, \emph{quantifying} these different PG histories empirically remains an active topic of investigation \citep{vinuesa_revisiting_2017, virgilio2025pressure, gomez2025linear, mahajan2025}.
However, PG histories are (broadly) described by the variations in $\beta(Re)$, or $\beta(x)$, observed upstream of the measurement location (\egg the accumulation of, and/or the rate of change of, $\beta$). 
The present experimental approach does not permit quantification of $\beta(\ret)$ or $\beta(x)$ with sufficient streamwise resolution; instead, we measure streamwise profiles of the pressure coefficient,
\begin{equation}
    C_{P}(x)=1- \frac{U_{\infty}^{2}(x)}{U_{\infty}^{2}(0)}, 
    \label{eq:cp}
\end{equation}
\noindent where $U_{\infty}(x)$ is the freestream velocity and $x=0$ denotes the wind tunnel test section inlet. 
The PG history can then be expressed as $\beta(x) = \frac{\delta^{*}}{C_{f}} \frac{dC_P}{dx} (1-C_{P})^{-1}$, where $C_{f}$ is the local friction coefficient. 
For most APG TBLs (especially the high-Reynolds-number, low-to-moderate APG TBLs which are relevant to this study), $\frac{\delta^{*}}{C_{f}}$ and $Re_{\tau}$ are expected to increase monotonically with $x$. 
As such, in the present study, any differences in $C_P(x)$ upstream of a given measurement location would directly result in (qualitative) differences in streamwise PG history (figure~\ref{fig1}b). 
It is also noted here that other upstream perturbations, such as discrete changes in wall-roughness or tripping conditions, also influence TBL development and are generally classified as \emph{history} effects \citep{hanson2016development}; however, the present study maintains hydrodynamically smooth walls and matched tripping conditions across all cases. 
Accordingly, the discussion of \emph{history} effects in this study is limited exclusively to differences in streamwise PG history.

\subsection{Present approach} 
\label{Approach}

\begin{figure}
  \captionsetup{width=1.00\linewidth}
   \begin{center}
    \includegraphics[width=1.00\textwidth]{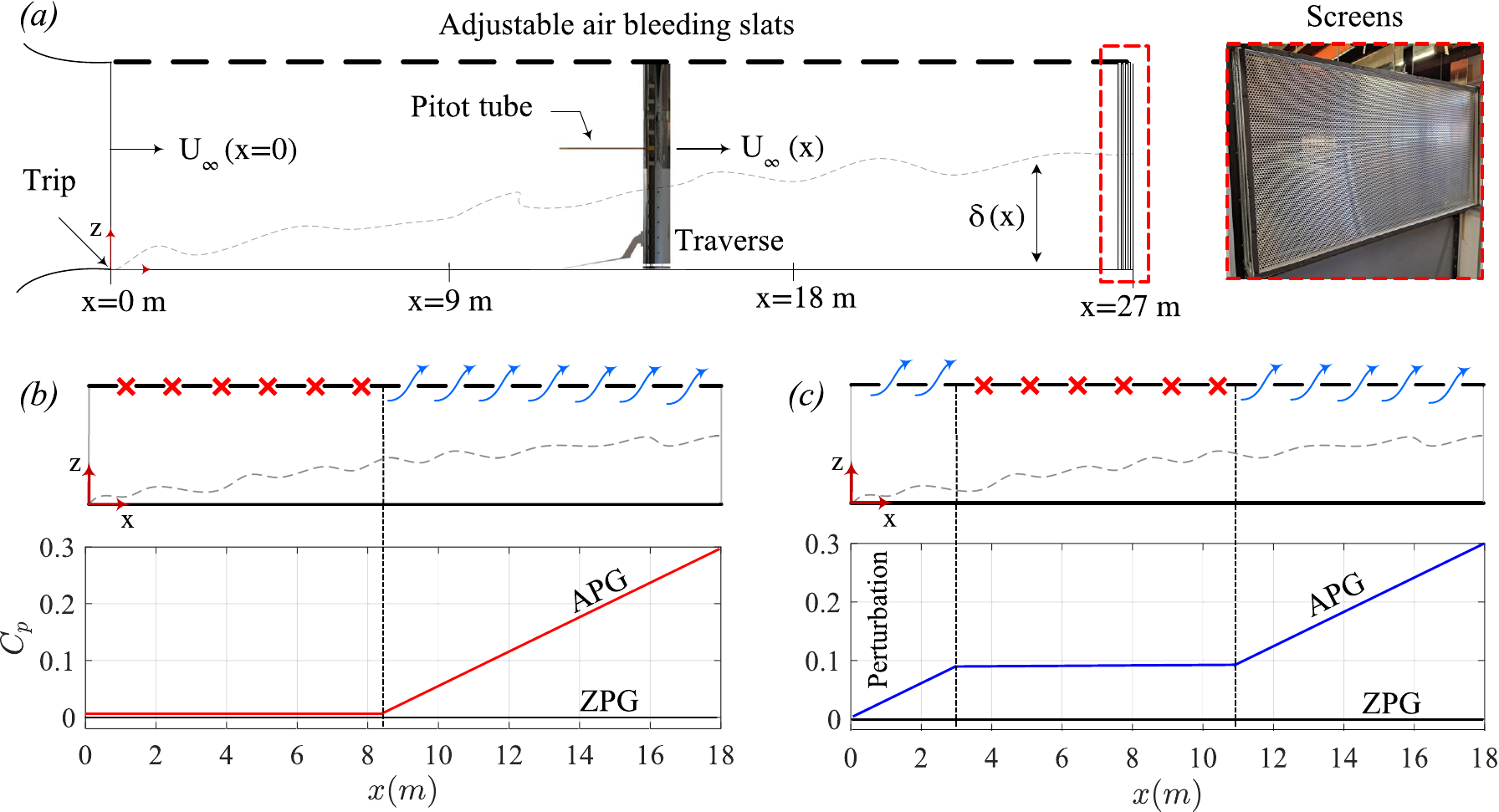}
   \end{center}
   \caption{(\textit{a}) Schematic of the experimental setup established in \citet{deshpande}, to generate high-$\ret$ APG TBLs with controlled upstream pressure-gradient histories. Conceptual profile of $C_{P}(x)$ along the test section to investigate a ZPG and APG TBLs with (\textit{b}) minimal PG history or (\textit{c}) with a controlled upstream PG perturbation, following the framework of figure~\ref{fig1}(b).}
   \label{fig2}
\end{figure}

\az{Figure \ref{fig1}(a) reveals a clear gap in the literature concerning the investigation of high-Reynolds-numbers TBLs under low-to-moderate adverse pressure gradients. 
Although datasets such as those of \citet{romero_properties_2022} cover a part of this parameter space, the influence of a unique PG history (due to a ramp) remains unquantified and coupled with the local APG. 
The present study addresses this gap by leveraging a unique large-scale experimental facility \citep{marusic_evolution_2015, deshpande}, and a systematic methodology, to generate high-$\ret$ moderate-APG TBLs with well-controlled upstream PG profiles (figure~\ref{fig2}), consistent with the conceptual framework illustrated in figure~\ref{fig1}.} 

First, this methodology is used to produce TBLs with matched $\ret$ and $\beta$ downstream, but with distinct upstream PG histories. 
This approach enables the decoupling of PG history effects from those of the local flow conditions on APG TBL development. 
In contrast to previous experiments, this methodology facilitates a controlled and systematic investigation of PG history effects, characterized by relatively weak upstream perturbations (low $|dC_{P}/dx|$), extended relaxation and development regions (as shown in figure \ref{fig1}c), and well-matched and controlled local flow conditions downstream. 
\az{The premise underlying this approach is that if differences in APG TBL statistics are detectable under weakly varying PG histories, it is reasonable to expect that similar effects will persist, and potentially strengthen, under stronger PG perturbations and/or shorter relaxation and development distances. 
As such, this systematic analysis enables more conclusive insights into the scaling of APG TBL statistics within the parameter space described above, which are also qualitatively relevant to other experiments, although we do not aim to directly extrapolate these quantitative effects to other parameter spaces.}
Additionally, the present comparisons between APG TBLs with different PG histories will provide insight into outstanding questions, such as the relaxation and development lengths required to damp upstream PG perturbations in high-$\ret$ experiments, the extent of TBL regions influenced by PG history, and related effects.

Finally, the same methodology will be applied to generate high-$\ret$ TBLs with varying low-to-moderate local APGs over an extended downstream fetch, while prescribing strictly controlled ZPG conditions upstream to minimize PG history effects, thereby addressing limitations in previous studies. 
The high-$\ret$ conditions, low-to-moderate APGs, and minimal PG history achieved here ensure a sufficiently broad overlap region extent to unambiguously investigate the influence of local APGs on the classical log-law. 
For this purpose, we will also incorporate oil-film interferometry (OFI) to obtain independent friction velocity measurements, which are crucial for the investigation of normalised turbulence statistics. 
\az{These results will also form the basis for Part~2, where new physical insights into the roles of $\ret$, local APGs, and PG history in TBL statistics will be combined with the framework of \citet{nickels} and a broad range of APG TBL datasets available in the literature to reconstruct a robust composite profile for the mean streamwise velocity.}

\begin{table}
  \begin{center}
  \renewcommand{\arraystretch}{1.0}
  \setlength{\tabcolsep}{5pt}
\def~{\hphantom{0}}
    \begin{tabular}{l c c c c c c c c c}
        \hline
        Case & $x$ & $\beta$ & $\mathrm{Re}_\tau$ & $\mathrm{Re}_\theta$ & $u_\tau$ & $\delta_{99}$  & $\delta$  & $\delta^\ast$  & Symbol \\
        & (m)& & & & & (mm) & (mm) & (mm) &\\
        \hline
        \multicolumn{6}{l}{\textit{From figure~\ref{fig3}(a)}}\\
        \\
        ZPG & 5.1 & 0 & 4090 & 8970 & 0.50 & 100.0 & 124.4 & 13.2 & \tikz\draw[draw=k1,line width=0.4mm]rectangle(6pt,6pt); \\
        Perturbed & 5.1 & 0 & 4060 & 9550 & 0.47 & 107.8 & 132.8 & 15.1 & \tikz\draw[draw=b5,fill=b1,line width=0.4mm]rectangle(6pt,6pt); \\
        \\
        ZPG & 8.5 & 0 & 5600 & 12840 & 0.48 & 139.0 & 182.9 & 19.4 & \tikz\draw[draw=k2,line width=0.4mm]rectangle(6pt,6pt);\\
        Perturbed & 8.5 & 0 & 5800 & 13580 & 0.46 & 151.6 & 200.1 & 21.6 & \tikz\draw[draw=b5,fill=b2,line width=0.4mm]rectangle(6pt,6pt); \\
        \\
        ZPG & 11.5 & 0 & 7000 & 16880 & 0.45 & 174.5 & 233.3 & 25.6 & \tikz\draw[draw=k3,line width=0.4mm]rectangle(6pt,6pt); \\
        Perturbed & 11.5 & 0 & 7400 & 17840 & 0.42 & 193.4 & 268.5 & 28.8 & \tikz\draw[draw=b5,fill=b3,line width=0.4mm]rectangle(6pt,6pt); \\
        \\
        ZPG & 12.8 & 0 & 8200 & 20360 & 0.48 & 205.8 & 273 & 29.6 & \tikz\draw[draw=k4,line width=0.4mm]rectangle(6pt,6pt); \\
        Perturbed & 12.8 & 0.64 & 8000 & 23200 & 0.41 & 234.4 & 296.5 & 37.8 & \tikz\draw[draw=b5,fill=b4,line width=0.4mm]rectangle(6pt,6pt); \\
        Ref. APG3 & 12.8 & 0.65 & 8000 & 23400 & 0.41 & 230.0 & 294.8 & 39.0 & \tikz\draw[draw=r5,fill=r4,line width=0.4mm]rectangle(6pt,6pt); \\
        \\
        ZPG & 17.4 & 0 & 10300 & 27656 & 0.43 & 276.1 & 370.1 & 41.0 & \tikz\draw[draw=k5,line width=0.4mm]rectangle(6pt,6pt); \\
        Perturbed & 17.4 & 1.47 & 10000 & 34400 & 0.36 & 348.8 & 433.9 & 65.2 & \tikz\draw[draw=b5,fill=b5,line width=0.4mm]rectangle(6pt,6pt); \\
        Ref. APG3 & 17.4 & 1.46 & 10100 & 34300 & 0.35 & 337.0 & 440.1 & 64.3 &\tikz\draw[draw=r5,fill=r5,line width=0.4mm]rectangle(6pt,6pt); \\
        \\
        \multicolumn{6}{l}{\textit{From figure~\ref{fig3}(b)}}\\
        \\
        ZPG & 8.5 & 0 & 4350 & 10428 & 0.33 & 148.0 & 203.2 & 23.3 & \tikz\draw[draw=k2,line width=0.4mm]circle(3.5pt); \\
        ZPG & 8.5 & 0 & 6300 & 14680 & 0.49 & 148.8 & 204.0 & 21.4 & \tikz\draw[draw=k2,line width=0.4mm]rectangle(6pt,6pt); \\
        APG1 & 8.5 & 0 & 4300 & 9921  & 0.35 & 146.0 & 193.2 & 21.4 & \tikz\draw[draw=g5,fill=g2,line width=0.4mm] circle (3.5pt); \\
        APG1 & 8.5 & 0 & 6500 & 16112 & 0.49 & 152.8 & 207.3 & 22.8 & \tikz\draw[draw=g5,fill=g2,line width=0.4mm]rectangle(6pt,6pt); \\
        APG3 & 8.5 & 0 & 4400 & 9760  & 0.34 & 140.5 & 195.4 & 20.9 & \tikz\draw[draw=r5,fill=r2,line width=0.4mm] circle (3.5pt); \\
        APG3 & 8.5 & 0 & 6200 & 15178 & 0.47 & 153.0 & 199.5 & 22.4 & \tikz\draw[draw=r5,fill=r2,line width=0.4mm]rectangle(6pt,6pt); \\   
        \\
        ZPG & 12.8 & 0 & 5800 & 14875 & 0.32 & 209.9 & 283.5 & 33.3 & \tikz\draw[draw=k4,line width=0.4mm] circle (3.5pt); \\
        ZPG & 12.8 & 0 & 8300 & 20360 & 0.48 & 205.8 & 273 & 29.6 & \tikz\draw[draw=k4,line width=0.4mm]rectangle(6pt,6pt); \\
        APG1 & 12.8 & 0.34 & 5700 & 15866 & 0.30 & 221.9 & 293.4 & 37.6 & \tikz\draw[draw=g5,fill=g4,line width=0.4mm] circle (3.5pt); \\
        APG1 & 12.8 & 0.34 & 8250 & 21325 & 0.44 & 216.3 & 284.3 & 32.7 & \tikz\draw[draw=g5,fill=g4,line width=0.4mm]rectangle(6pt,6pt); \\
        APG3 & 12.8 & 0.58 & 5750 & 17036 & 0.28 & 233.4 & 310.9 & 41.9 & \tikz\draw[draw=r5,fill=r4,line width=0.4mm] circle (3.5pt); \\
        APG3 & 12.8 & 0.57 & 8400 & 22543 & 0.43 & 224.7 & 305.4 & 36.1 & \tikz\draw[draw=r5,fill=r4,line width=0.4mm]rectangle(6pt,6pt); \\
        \\
        ZPG & 17.4 & 0 & 7300 & 20048 & 0.32 & 276.9 & 368.1 & 44.6 & \tikz\draw[draw=k5,line width=0.4mm] circle (3.5pt); \\
        ZPG & 17.4 & 0 & 10600 & 27656 & 0.43 & 276.1 & 370.1 & 41.0 & \tikz\draw[draw=k5,line width=0.4mm]rectangle(6pt,6pt); \\
        APG1 & 17.4 & 0.66 & 7000 & 22246 & 0.27 & 310.3 & 399.0 & 56.0 & \tikz\draw[draw=g5,fill=g5,line width=0.4mm] circle (3.5pt); \\
        APG1 & 17.4 & 0.67 & 10000 & 30749 & 0.39 & 300.4 & 387.0 & 50.9 & \tikz\draw[draw=g5,fill=g5,line width=0.4mm]rectangle(6pt,6pt); \\
        APG3 & 17.4 & 1.44 & 6800 & 24757 & 0.25 & 337.8 & 420.9 & 68.1 & \tikz\draw[draw=r5,fill=r5,line width=0.4mm] circle (3.5pt); \\
        APG3 & 17.4 & 1.46 & 9700 & 34248 & 0.36 & 332.4 & 419.3 & 63.7 & \tikz\draw[draw=r5,fill=r5,line width=0.4mm]rectangle(6pt,6pt); \\
        \hline
        
    \end{tabular}
    \caption{Characteristics of the experimental TBLs investigated in the present study.}
     \label{tab:1}
  \end{center}
\end{table}

\section{Experimental setup and methodology}
\label{ExpMethod}

\az{Experiments were conducted in the high-Reynolds-number boundary layer wind tunnel at the University of Melbourne. 
The test section has a cross-sectional area of $1.89 \times 0.92\ \rm{m}^{2}$ and a working length of 27~m (figure~\ref{fig2}a), enabling the development of physically thick TBLs with high friction Reynolds numbers toward the downstream end. 
Inlet flow conditioning ensures that freestream turbulence intensities remain below 0.4\% throughout the test section.
The test section ceiling is equipped with an array of adjustable air-bleed slots spanning the full width, distributed quasi-periodically in the streamwise direction. 
These slots permit controlled bleeding of air from the test section, driven by the internal static pressure which is regulated by altering the blockage at the test section outlet. 
For example, a nominal ZPG TBL is obtained by installing a single high-porosity mesh at the outlet and keeping all air-bleed slots fully open.
A detailed characterization of this setup for ZPG TBLs is provided by \citet{marusic_evolution_2015}. 
Following \citet{deshpande}, the test section static pressure can be increased by installing one or more low-porosity (higher-blockage) screens at the outlet. 
This produces nominally APG conditions where the air-bleed slots are open, and ZPG conditions where they are restricted or closed (figures~\ref{fig2}b,c). 
This approach enables the generation of well-controlled APG TBLs with minimal physical modifications to the test section, while remaining representative of practical conditions. 
Figures~\ref{fig2}(b,c) also illustrate how different PG histories, \ie\ streamwise variations in pressure coefficient, can be achieved by selectively opening and/or closing these air-bleed slots.}

\begin{figure}
  \captionsetup{width=1.00\linewidth}
   \begin{center}
    \includegraphics[width=1.00\textwidth]{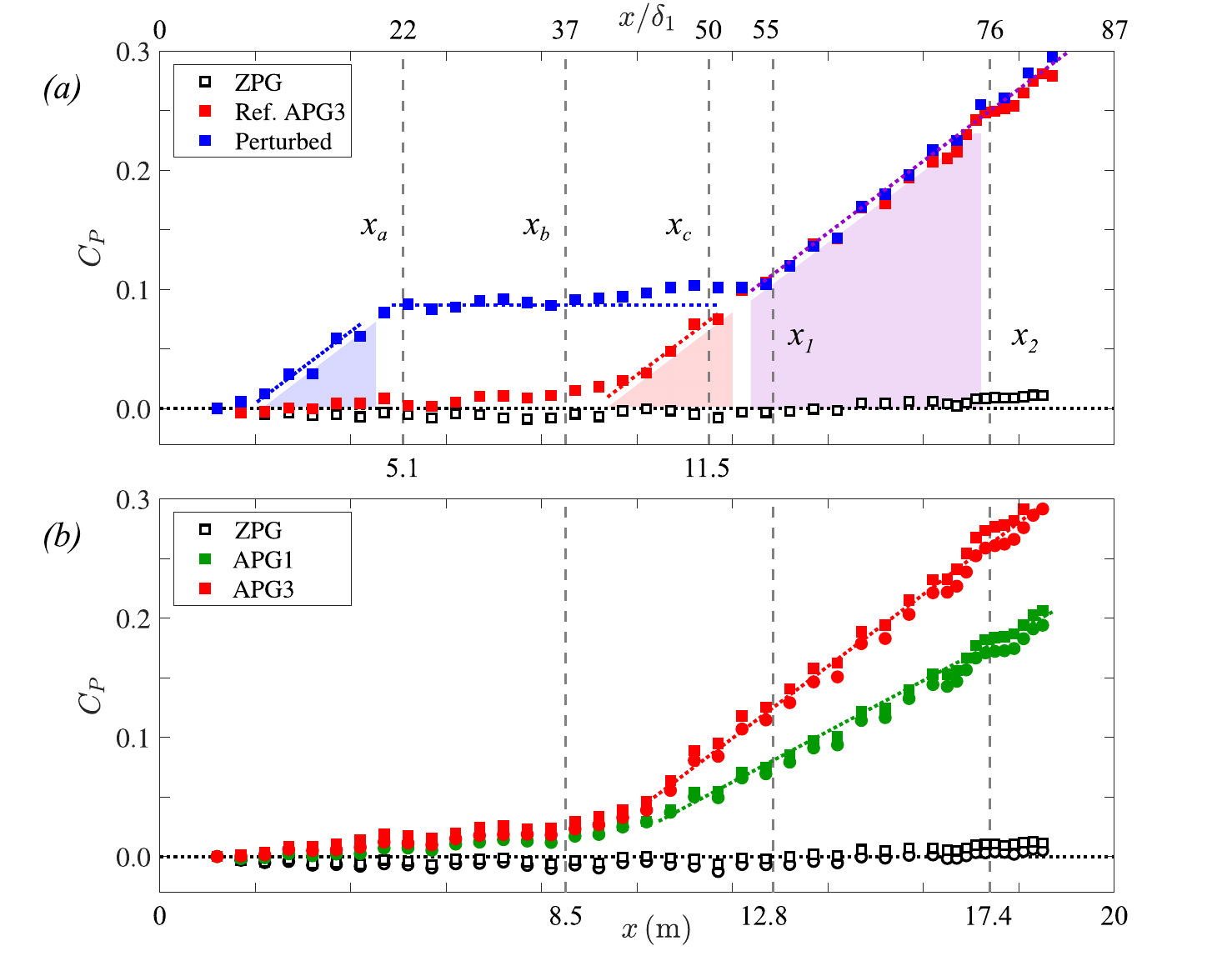}
   \end{center}
   \caption{$C_P(x)$ profiles measured for the (\textit{a}) ZPG, Ref. APG3 and Perturbed cases, and the (\textit{b}) ZPG, APG1 and APG3 cases (see table~\ref{tab:1} for exact flow conditions). Vertical dashed lines indicate the locations of streamwise velocity and OFI measurements. $\delta_1$ denotes the value of $\delta$ at $x_1$. \az{In the APG regions, $dC_P/dx$ is nominally 0.025 for the APG3 and Perturbed cases and 0.015 for the APG1 case.}}
   \label{fig3}
\end{figure}

\subsection{Imposing controlled pressure-gradient histories}
\label{Methodology}

Figure~\ref{fig3} presents the experimentally measured pressure coefficient profiles, $C_P(x)$, for various test section configurations, as listed in table~\ref{tab:1}. 
For each configuration, the freestream dynamic pressure was measured along the streamwise extent of the test section using a Pitot-static tube mounted at $z = 0.525$ m on the streamwise traverse system (figure~\ref{fig2}a), enabling measurement of the streamwise pressure coefficient profile.
The most basic configuration corresponds to nominal ZPG conditions throughout the test section, indicated by black symbols in figure~\ref{fig3}. 
Across all ZPG cases tested, variations in $U_{\infty}$ remained within 1\% along the entire test section fetch (resulting in $C_P(x)=0\pm0.012$). 
In figure~\ref{fig3} and table~\ref{tab:1}, square and circle symbols denote inlet unit Reynolds numbers of \( U_{\infty}(x=0)/\nu \approx 9 \times 10^5\ \mathrm{m}^{-1} \) and \( 6 \times 10^5\ \mathrm{m}^{-1} \), respectively. 
To generate different APG configurations, either one or three low-porosity screens were added to the outlet, in addition to the high-porosity mesh. 
These are denoted as APG1 (green symbols in figure~\ref{fig3}b) and APG3 (red symbols in figures~\ref{fig3}a,b). 
As shown in figure~\ref{fig3}(b), both the APG1 and APG3 configurations exhibit nominal ZPG conditions upstream of the APG development region (\( x < 9\ \mathrm{m} \)), consistent with the setup illustrated in figure~\ref{fig2}(b). 
These cases are referred to as having \emph{minimal upstream PG history} and serve as the reference cases for comparison against those with selectively imposed PG histories. 
The TBL statistics for the APG1 and APG3 cases were found to collapse with those from the ZPG case at \( x = 8.5\ \mathrm{m} \), confirming the absence of upstream PG history variations, consistent with the findings of \citet{deshpande}. 
An additional configuration, referred to as the \emph{Perturbed} case, was then designed to introduce a controlled upstream PG history which was distinct from the baseline APG cases. 
Specifically, this setup modifies the pattern of open/closed air-bleed slots in the upstream region, relative to the APG3 case (figures~\ref{fig2}b,c). 
The resulting \( C_P(x) \) profile is shown by blue symbols in figure~\ref{fig3}(a). 
For this case, the inlet and downstream flow conditions were matched to the APG3 reference configuration (Ref.\ APG3), but notable differences in \( C_P(x) \) appear upstream (\( x < 12\ \mathrm{m} \)) due to the imposed PG perturbation. 
\az{Appendix~\ref{apxC}, contains additional details about the exact configurations of open and closed slats required to generate the Perturbed and Ref. APG3 cases.}
The Perturbed case was also designed to include a long relaxation region (\( 5 < x < 12\ \mathrm{m} \)), with nominally ZPG conditions, following the initial PG perturbation. 
This enables investigation of the persistence of PG history effects induced by the upstream PG perturbation. 
Further downstream, the facility permits sustained APG development over a long streamwise distance, enabling a detailed comparison of the differences in development between the Perturbed and Ref. APG3 cases under locally matched flow conditions (\( x > 12\ \mathrm{m} \)).

\subsection{Streamwise velocity and friction velocity measurements}
\label{Measurements}

Five measurement stations were chosen, spanning a significant streamwise fetch, where both the turbulence statistics and friction velocity could be measured reliably. 
\az{Specifically, the stations $x_a$ and $x_1$ were chosen because they are expected to exhibit the highest contrast in turbulence statistics between the Perturbed and reference cases. 
The stations $x_b$, $x_c$, and $x_2$ are then used to assess the recovery of the Perturbed case towards the reference cases.}

\az{Streamwise velocities were measured at several wall-normal locations using a single hot-wire made in-house from Wollaston wire mounted on a Dantec 55P15 probe. 
The wire was etched to a sensing diameter \( d = 2.5\,\mu\mathrm{m} \) with fixed aspect ratio \( l/d = 200 \), giving a nominal length \( l = 0.5\,\mathrm{mm} \) and resulting in viscous-scaled lengths \( l^+ = 11{-}16 \) across all cases. 
To resolve the energetic hierarchy of scales in the TBL, the sampling time \( T_S \) was chosen to satisfy \( T_S U_{\infty}/\delta \ge 20{,}000 \). 
The sampling frequency was fixed at 50 kHz and the signal was low-pass filtered at 25 kHz. 
The resulting viscous-scaled sampling intervals \( t_S^+ = \ut^2 /(f_S \nu) \) ranged from 0.08 to 0.31, sufficient to resolve the relevant small-scale turbulence. 
Further details of the experimental apparatus and calibration methods are available in \citep{deshpande}.} 
OFI was used at each streamwise measurement location (except $x=x_c$) to obtain an accurate and independent estimate of \( \ut \).
Further details of the OFI methodology implemented in the present APG TBL experiments are provided in \citet{marusic2024turbulent}. 
Taking into account uncertainties associated with oil viscosity calibration, interference pattern resolution, and potential contamination of the oil droplets, the estimated uncertainty in the measured friction velocity was approximately $\pm 1$\%. 

\subsection{Summary of present experimental dataset}
\label{dataset}

Key characteristics for the TBLs investigated here, obtained from these hot-wire, OFI, and $C_P$ measurements at each measurement location, along with their respective symbols, are presented in table~\ref{tab:1}, for reference. 
The recent approach of \citet{lozier2025defining} was employed to determine the TBL thickness ($\delta$), which is compared with the classical $\delta_{99}$, as defined by \citet{vinuesa2016delta}, in table~\ref {tab:1}. 
The capability of the proposed methodology for creating controlled PG histories can now be demonstrated quantitatively using the characteristics tabulated in table~\ref{tab:1}. 
First, for the cases in figure~\ref{fig3}(a), $\ret(x)$ is measured to be nominally matched between the Perturbed and ZPG cases in the upstream section ($x < 12$ m). 
This is an important condition for investigating the persistence of any PG history effects observed within the relaxation region that may result from the upstream APG perturbation. 
In the downstream section ($x > 12$ m) where $C_P$ and $dC_P/dx$ are matched for the Perturbed and Ref.\ APG3 cases (figure~\ref{fig3}a), $\ret(x)$ and $\beta(x)$ are also confirmed to be nominally matched (table~\ref{tab:1}). 
This is the critical condition for decoupling the local and upstream PG effects \citep{bobke2017history, mahajan2025}, and quantifying the influence of upstream PG perturbations on the APG TBL development in the downstream development region. 
Finally, in figure~\ref{fig3}(b), although the local PG (i.e. $\beta$) varies, the Reynolds number is found to be nearly identical between the ZPG, APG1 and APG3 cases with matched inlet conditions, as shown in table~\ref{tab:1}. 
This ensures that the effects of the local pressure gradient on TBL statistics and scaling laws can be rigorously evaluated without the influence of varying Reynolds numbers and/or PG history effects.

\section{PG history effects in high-Reynolds-number APG TBLs}
\label{Results}

\subsection{Streamwise mean velocity}  
\label{MeanVel}

\begin{figure}
  \captionsetup{width=1.00\linewidth}
   \begin{center}
    \includegraphics[width=1.00\textwidth]{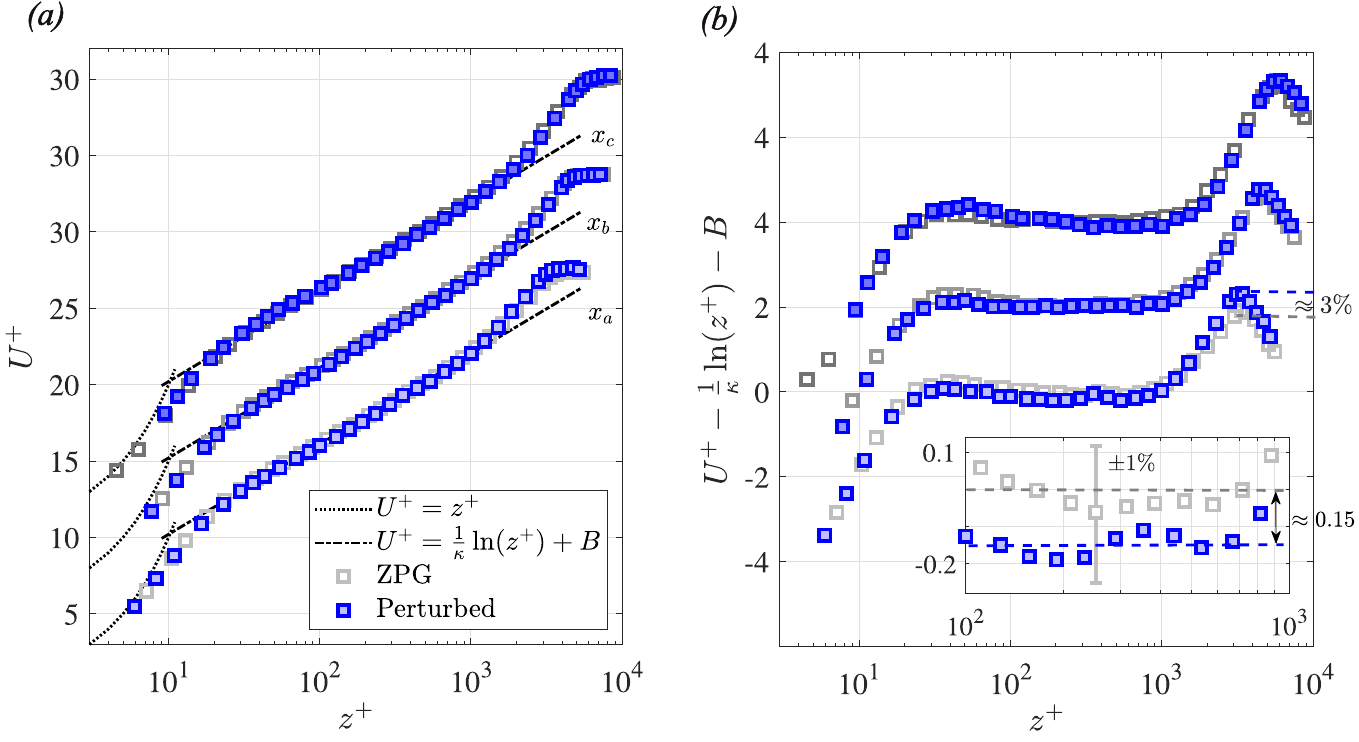}
   \end{center}
   \caption{(\textit{a}) Profiles of mean streamwise velocity, and (\textit{b}) deviations from the classical log-law for the ZPG and Perturbed cases in the relaxation region. Both cases have locally matched flow conditions, but unique PG histories. Note that the profiles for $x_b$ and $x_c$ are vertically offset.}
   \label{fig4}
\end{figure}

We first examine viscous-scaled mean streamwise velocity profiles ($\up$) measured in the relaxation region, \ie at $x_a$, $x_b$ and $x_c$ (figure \ref{fig3}a), which are shown in figure~\ref{fig4}(a). 
The purely ZPG case is used here as a reference. 
Although the flow conditions of the Perturbed case are locally matched to the ZPG case in the relaxation region (table~\ref{tab:1}), it has experienced an upstream APG perturbation (figure~\ref{fig3}a) and the resulting differences in their mean velocity profiles can be seen at $x_a$. 
Specifically, the Perturbed mean velocity profile ($\up$) in the wake region lies above the ZPG profile and the Perturbed profile in the overlap region is slightly below the ZPG profile at $x_a$ (figure~\ref{fig3}a). 
This is made clearer by subtracting the classical log-law \eqref{eq:loglaw} from these measured mean velocity profiles ($\up-\up_{log}$), which reveals a systematic deviation from canonical scaling behaviour for the Perturbed profile in figure~\ref{fig4}(b). 

It is also noted that the mean velocity profile in the inner region remains largely unaffected by the upstream APG perturbation, consistent with the findings of \citet{bobke2017history} and \citet{pozuelo}. 

Further downstream, the Perturbed mean velocity profiles collapse with the reference ZPG profiles at $x_b$ and $x_c$ (figure~\ref{fig4}a). 
This collapse is further confirmed by figure~\ref{fig4}(b), where both cases show good agreement with the classical log-law. 
However, differences in characteristics such as the TBL thickness ($\delta$, see table~\ref{tab:1}), and its growth rate ($d\delta/dx$), still persist and are attributed to the upstream APG perturbation. 
For example, across the full relaxation region (\ie between $x_a$ and $x_c$ where $x_c - x_a \approx 28\delta_1$) the growth rate of the Perturbed TBL is 25\% greater than the reference ZPG TBL. 
\az{This effect is attributed to the large-scale motions, which are energized by the upstream APG perturbation (see figure~\ref{fig6}c), and enhance momentum transport in the TBL. 
Both observations are significant: the mean velocity profile does recover ZPG scaling, suggesting that the effect of upstream PG perturbations (like those introduced by ramps) on mean turbulence statistics can be minimized with a sufficiently long relaxation length \citep[][see figure~\ref{fig1}]{romero_properties_2022, knopp2021experimental}, but the boundary-layer thickness and growth are not equivalent, making careful comparisons challenging, even after a reasonably long relaxation length.} 
These results also motivate further investigation into the quantification of PG history effects, specifically methods/parameters which can account for the differences in behaviour observed between $x_a$ and $x_b$ (or $x_c$).

\begin{figure}
  \captionsetup{width=1.00\linewidth}
   \begin{center}
    \includegraphics[width=1.00\textwidth]{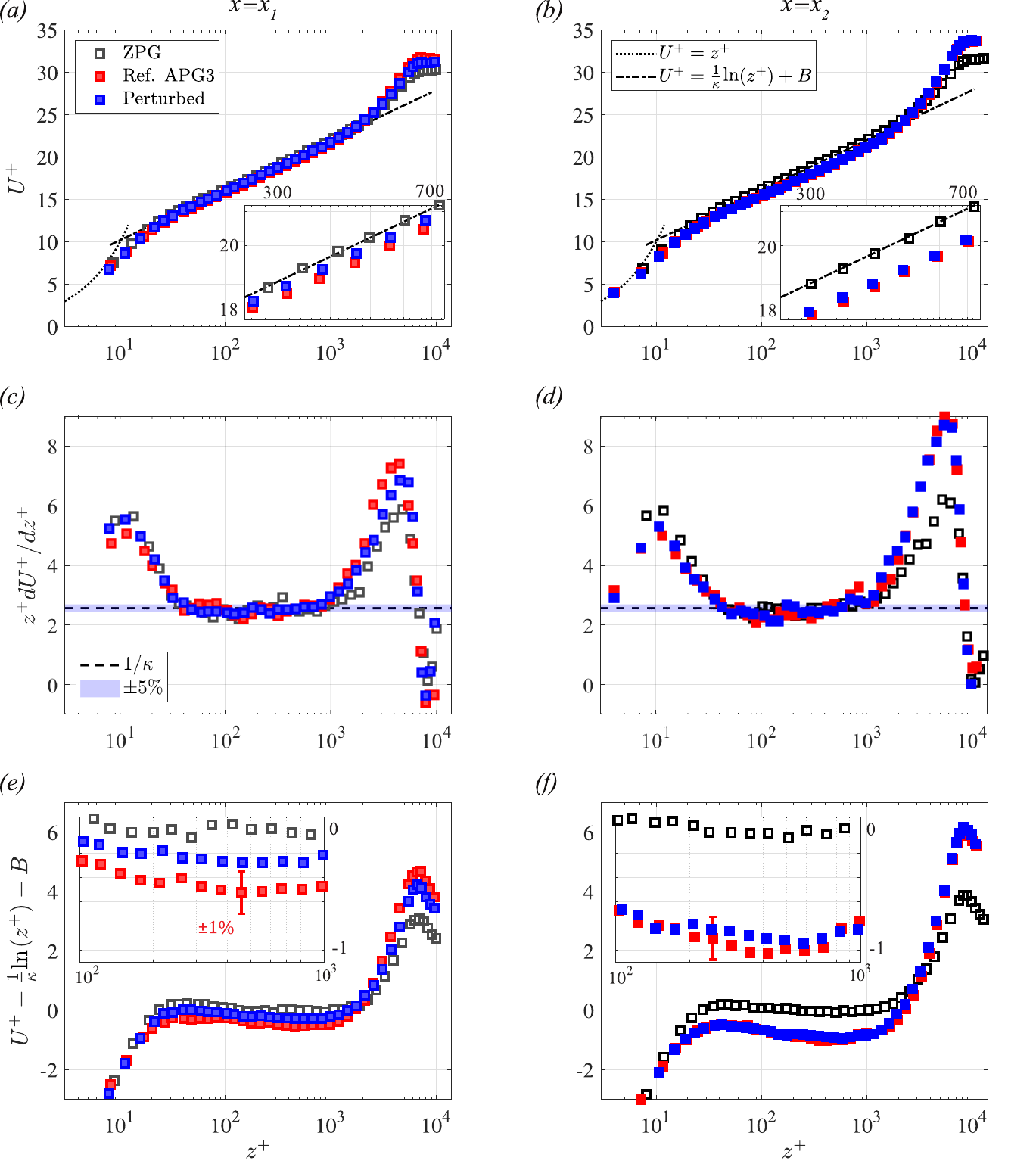}
   \end{center}
   \caption{(\textit{a,b}) Profiles of mean streamwise velocity, (\textit{c,d}) indicator function, and (\textit{e,f}) deviations from the classical log-law for the ZPG, Ref. APG3 and Perturbed cases in the development region at (\textit{left}) $x=x_1$ and (\textit{right}) $x=x_2$. All \az{APG} cases have locally matched flow conditions, but unique PG histories. \az{The ZPG cases have locally matched $\ret$, for reference.}}
   \label{fig5}
\end{figure}

The mean velocity profiles measured in the development region, \ie at $x_1$ and $x_2$ (figure \ref{fig3}a), are shown in figures~\ref{fig5}(a,b). 
Here, the Perturbed and Ref.\ APG3 cases have nominally matched values of $\ret$ and $\beta$ (table~\ref{tab:1}), and the ZPG case has matched $\ret$, for reference. 
Since the mean velocity profiles for all cases collapse at $x_b$ (figure~\ref{fig4}a), any differences observed between the Perturbed and Ref.\ APG3 profiles at $x_1$ and/or $x_2$ can be attributed solely to the differences in PG history between $x_b$ and $x_1$. 
While, the mean velocity profiles very near the wall ($\zp\leq 20$) appear largely unaffected by either the local or upstream PGs at both $x_1$ and $x_2$ \citep{pozuelo}, the effect of the local APG ($\beta$) can be observed at $x_2$ where the Ref.\ APG3 and Perturbed profiles collapse and deviate strongly from the ZPG profile in the wake and overlap regions \citep{monty2011,deshpande}. 
However, at $x_1$, the influence of PG history is also apparent; although both the Perturbed and Ref. APG3 profiles deviate from the ZPG case, they do not collapse with one another, consistent with the observations of \citet{bobke2017history}. 
These results further motivate the development of methods/frameworks for quantifying PG history effects that can account for the differences in behaviour observed between $x_1$ and $x_2$; one such framework, motivated by the above observations, is proposed in Part~2. 

The influence of local and upstream PGs on the classical log-law is further investigated by examining the indicator function ($\zp d\up /d\zp$) of each mean velocity profile, as shown in figures~\ref{fig5}(c,d). 
At both measurement locations, and for all PG cases, a clear plateau is observed in the overlap region ($10^2\lesssim \zp \lesssim10^3$), confirming the existence of logarithmic scaling and the sufficient separation between the inner and wake regions. 
Additionally, within experimental uncertainty, no systematic changes are observed in the von K\'arm\'an coefficient ($\kappa$) due to either local or upstream PGs. 
The differences in mean velocity profiles in the wake region confirm that the wake parameter ($\Pi$) systematically increases with $\beta$ \citep[figure~\ref{fig5}d, ][]{monty2011}, as well as by PG history \citep[figure~\ref{fig5}c, ][]{bobke2017history}. 
Finally, the classical log-law \eqref{eq:loglaw} was subtracted from the measured mean velocity profiles in figures~\ref{fig5}(e,f), in a manner analogous to figure~\ref{fig4}(b). 
In the overlap region, a systematic and significant decrease in the magnitude of the additive coefficient $B$ is observed with increasing $\beta$, especially at $x_2$. 
However, PG history is also seen to affect the value of $B$ at $x_1$, where the Ref.\ APG3 and Perturbed cases do not collapse. 
\az{It is also noted that the mean velocity profile of the Perturbed case at $x_1$ also does \textit{not} match the ZPG case, as shown in figure~\ref{fig5}(a,c,e). 
Clear differences appear in the wake and overlap regions, which are beyond experimental uncertainty, indicating a clear influence of the local APG even with a relatively short exposure to these APG conditions (\ie $x_1$ is sufficiently downstream of the region with nominally ZPG conditions).} 

\subsection{Streamwise normal stress and premultiplied energy spectra}
\label{StressSpectra}

We now examine the influence of local and upstream PGs on second-order turbulence statistics and the hierarchy of energetic turbulent motions in the TBL. 
To this end, it is useful to contextualise the role of large-scale motions (LSMs) in high-$\ret$ TBLs (e.g. motions associated with attached-eddies and/or superstructures). 
These large-scale motions constitute the primary energy-containing motions in the overlap and wake regions of these high-$\ret$ TBLs, and their contributions become increasingly significant with higher Reynolds numbers \citep{marusic_logarithmic_2013, marusic_evolution_2015}. 
In the context of APG flows, it has been established that the magnitude and extent of these LSMs and/or superstructures, in the overlap and wake regions, are influenced by local PGs \citep{deshpande}. 
Recent studies \citep{bobke2017history, pozuelo} have also demonstrated that upstream PGs can leave a persistent footprint on the LSMs in the wake region of low-Reynolds-number TBLs. 
Hence, it is important to decouple the influences of these local and upstream PGs on the hierarchy of inertia-dominated turbulent motions.
The present high-$\ret$ experiments are critical to achieve this objective as they offer sufficient separation between the viscous and inertia-dominated regions of the TBL to facilitate detailed examination of these influences.  

\begin{figure}
  \captionsetup{width=1.00\linewidth}
   \begin{center}
    \includegraphics[width=1.00\textwidth]{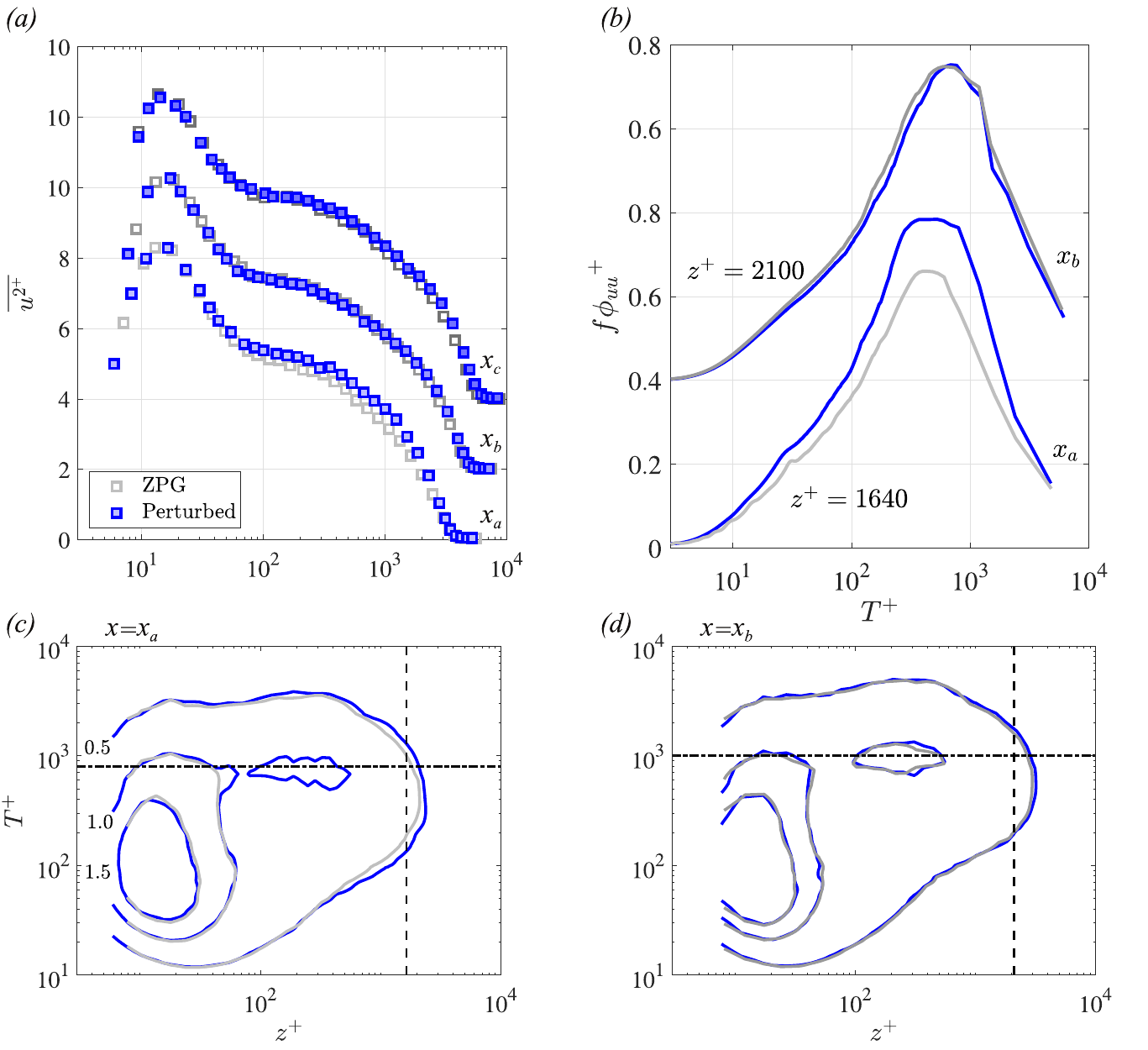}
   \end{center}
   \caption{(\textit{a}) Profiles of streamwise normal stress for the ZPG and Perturbed cases in the relaxation region. Premultiplied energy spectra ($f\phi_{uu}^+$) at (\textit{c}) $x=x_a$ and (\textit{d}) $x=x_b$. Vertical dashed lines indicate $z^{+}=0.4\ret$. Horizontal dot-dashed lines denote $T^+=4.8\delta^{+}/U_{\infty}^{+}$. (\textit{b}) Premultiplied spectra at a single wall-normal location corresponding to the vertical dashed lines in (\textit{c,d}).
   Note that the profiles for $x_b$ and $x_c$ are vertically offset in (a) and (b).}
   \label{fig6}
\end{figure}

The streamwise normal stress profiles ($\overline{u^{+2}}$) measured in the relaxation region are presented in figure~\ref{fig6}(a), corresponding to the same cases presented in figure~\ref{fig4}(a). 
To investigate the distribution of turbulent energy across different scales, the premultiplied spectra of streamwise velocity fluctuations ($f\phi_{uu}^+$) have been computed and are plotted as a function of the viscous time scale ($T^+=\ut^2/ f\nu$) and wall-normal location ($\zp$) in figures~\ref{fig6}(c,d) corresponding to the streamwise locations $x_a$ and $x_b$, respectively. 
Some differences in the streamwise normal stress profiles (figure~\ref{fig6}a) and premultiplied spectra (figure~\ref{fig6}c) are observed in the wake region ($\zp\gtrsim100$) at $x_a$ between the ZPG and Perturbed cases, which can be attributed to differences in their respective PG histories upstream of the measurement location (figures~\ref{fig3}a,c). 
In figures~\ref{fig6}(c,d), the horizontal dash-dotted lines denote the characteristic, Reynolds-number-dependent time scale associated with LSMs in canonical TBLs. 
At \( x_a \), a secondary spectral peak appears for the Perturbed case (figure~\ref{fig6}c), located slightly below this line, i.e. at smaller time scales, highlighting the influence of the upstream APG perturbation \citep{deshpande}. 
This spectral feature also corresponds with elevated wake region turbulence intensities in the Perturbed normal stress profile, and reinforces the conclusion that the upstream PG perturbation leaves a measurable footprint at this location. 
Further downstream, however, at \( x_b \), both the normal stress profiles (figure~\ref{fig6}a) and premultiplied spectra (figure~\ref{fig6}d) for the ZPG and Perturbed cases collapse, indicating recovery from the upstream PG perturbation. 
This collapse also persists at \( x_c \); however, analysis at that location is omitted here for brevity, citing consistency with the trends noted in the mean velocity profiles.

To further investigate the influence of the upstream PG perturbation on the hierarchy of turbulent scales, spectra corresponding to the wake region ($\zp=0.4\ret$; vertical dashed lines in figures~\ref{fig6}c,d) are plotted in figure~\ref{fig6}(b) for $x_a$ and $x_b$. 
Here, $0.4 \ret$ is approximately equivalent to $0.5 \delta_{99}^+$, for reference, and this particular wall-normal location was chosen because it has previously been noted for being strongly influenced by local PGs \citep{deshpande}. 
While the spectra at this wall-normal location collapse at $x_b$, the influence of the upstream APG perturbation on the Perturbed case can be seen across a broad range of time scales at $x_a$. 
However, the `inner-peak'  of $\overline{u^{+2}}$ and the associated characteristic wall-normal location/scale in the spectrograms ($\zp=15$, $T^+=100$) remain unaffected by the upstream APG perturbation at $x_a$ (figures~\ref{fig6}a,c), which is consistent with \citet{bobke2017history} and \citet{pozuelo}. 
These observations reaffirm that small-scale motions in the inner region may respond more rapidly to changes in the local PG conditions, \ie they are less sensitive to PG history, than the LSMs and small-scale turbulence in the wake region \citep{marusic_evolution_2015, sanmiguel_vila_experimental_2020, gungor2024turbulent, mahajan2025}. 

\begin{figure}
  \captionsetup{width=1.00\linewidth}
   \begin{center}
    \includegraphics[width=1.00\textwidth]{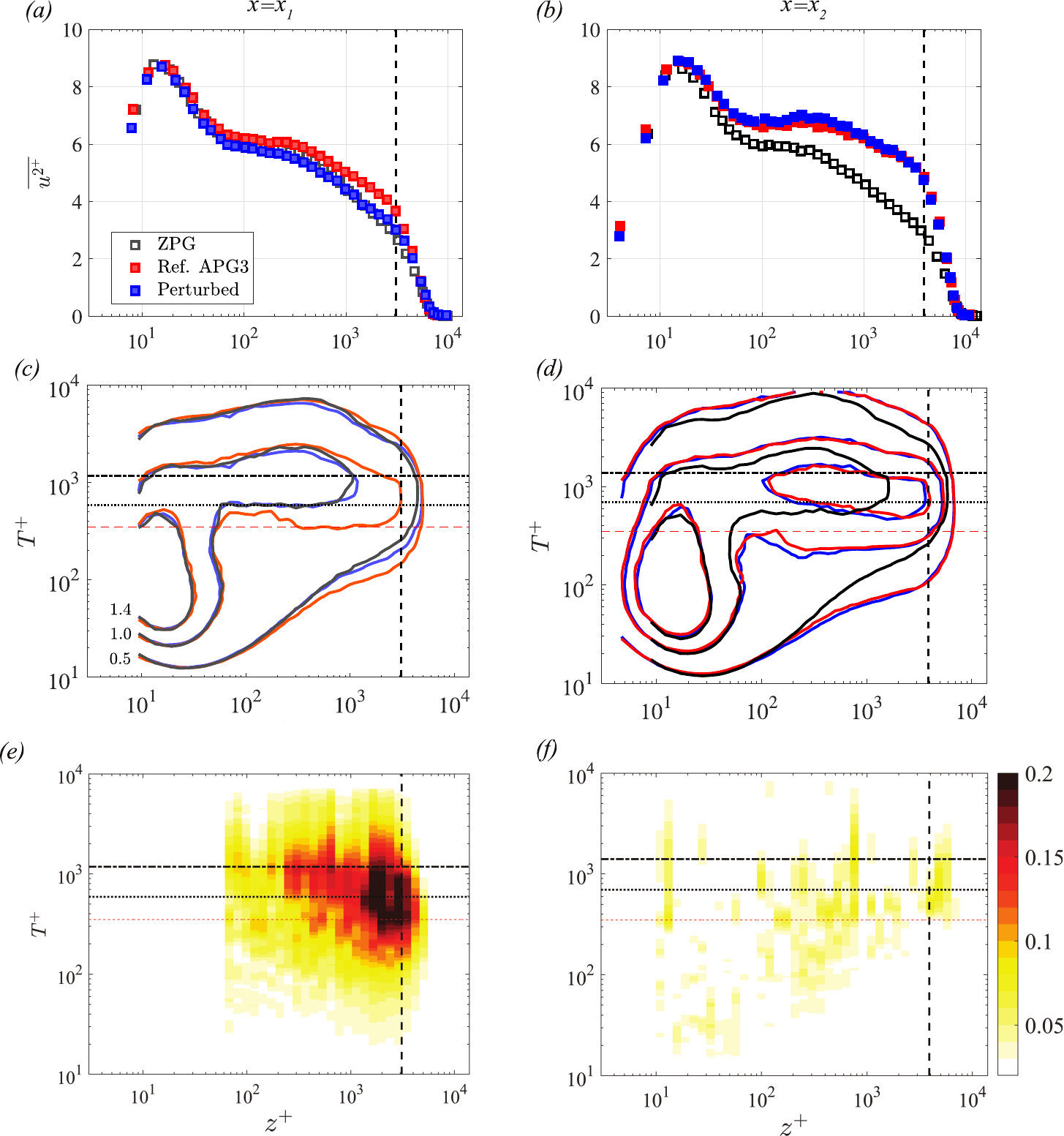}
   \end{center}
   \caption{(\textit{a,b}) Profiles of streamwise normal stress and (\textit{c,d}) premultiplied energy spectra ($f\phi_{uu}^+$) for the ZPG, Ref. APG3 and Perturbed cases in the development region at (\textit{left}) $x=x_1$ and (\textit{right}) $x=x_2$. (\textit{e,f}) Differences between the Ref. APG3 and Perturbed spectra ($f\phi_{uu,\rm{Ref.\,APG3}}^+-f\phi_{uu,\rm{Perturbed}}^+$). Horizontal dot-dashed lines denote $T^+=4.8\delta^{+}/U_{\infty}^{+}$, horizontal dotted lines denote $T^+=2.4\delta^{+}/U_{\infty}^{+}$, \az{ and the horizontal red dashed line represents $T^+=350$.}}
   \label{fig7}
\end{figure}

We now investigate the influence of local and upstream PGs farther downstream in the development region, \ie at $x_1$ and $x_2$. 
Consistent with observations from the relaxation region, $\overline{u^{+2}}$ profiles in the inner region collapse across all cases (figures~\ref{fig7}a,b). 
However, the effects of the local and upstream PGs are evident in the wake region where LSMs are dominant (vertical dashed lines again indicate $\zp=0.4\ret$).
At $x_2$, the Perturbed and Ref.\ APG3 cases collapse, but both deviate significantly from the ZPG case across a broad wall-normal range, including the overlap and wake regions (figure~\ref{fig7}b), highlighting the expected effect of local APGs \citep{deshpande}. 
In contrast, at $x_1$, while the Ref.\ APG3 case clearly departs from the ZPG case in the wake region (figure~\ref{fig7}a), the Perturbed case does not match the Ref.\ APG3 case, reflecting differences in their PG histories upstream of the measurement location. 
Specifically, the magnitude of the normal stress for the Perturbed case only deviates from the ZPG case around $\zp=0.4\ret$, likely owing to its predominantly ZPG history upstream of $x_1$ (figure~\ref{fig3}). 

To further examine the impact of local and upstream PGs on the hierarchy of energetic motions, the corresponding premultiplied energy spectrograms are shown in figures~\ref{fig7}(c,d). 
Both the ZPG and Ref. APG3 cases display typical distributions of energetic turbulent motions; for instance, the outer spectral peak in the ZPG case aligns with the horizontal dash-dotted line \citep[$T^+= 4.8\delta^+/U_{\infty}^+ \approx 6\delta_{99}^+/U_{\infty}^+$;][]{marusic_evolution_2015}, while the outer peak in the Ref. APG3 case (reflecting the effect of the local APG) aligns with the horizontal dotted line \citep[$T^+ = 2.4\delta^+/U_{\infty}^+ \approx 3\delta_{99}^+/U_{\infty}^+$;][]{deshpande}.
Building on the observations from the streamwise normal stress, the Perturbed case at $x_1$ falls between the ZPG and Ref. APG3 cases due to its upstream PG history, whereas at $x_2$ it collapses with the Ref. APG3 case. 
This confirms that the influences of PG variations upstream of $x_1$ do not persist at $x_2$.  
\az{It is important to note that, at $x_1$, the Perturbed case \emph{is} already responding to the local APG conditions, i.e. not purely ZPG, and the differences observed in the normal stress and mean velocity around $\zp = 0.4\ret$ are connected to the energization of a broad range of scales.}

To further decouple the influence of PG history, the Ref. APG3 spectra were subtracted from the Perturbed spectra, \ie $(f\phi_{uu}^+)_{\rm{Ref\,.APG3}}-(f\phi_{uu}^+)_{\rm{Perturbed}}$, and the resulting differences are shown in figures~\ref{fig7}(e,f). 
At $x_1$, the most prominent differences appear in the wake region, centred around the expected location and scale of APG effects ($\zp = 0.4\ret$, $T^+ = 2.4\delta$), demonstrating how the Perturbed case `lags' behind the Ref. APG3 case in APG development due to differences in their upstream PG histories. 
Smaller differences are also observed in the logarithmic region at $x_1$ at larger time scales, supporting the idea that larger-scale motions take longer to respond to changes in PG conditions, that is, they retain some measurable `memory' of the upstream PG history. 
At $x_2$, however, no significant differences remain in the subtracted spectra, confirming that the two cases are locally matched and that PG history-related effects are no longer significant. 
 
\begin{figure}
  \captionsetup{width=1.00\linewidth}
   \begin{center}
    \includegraphics[width=0.9\textwidth]{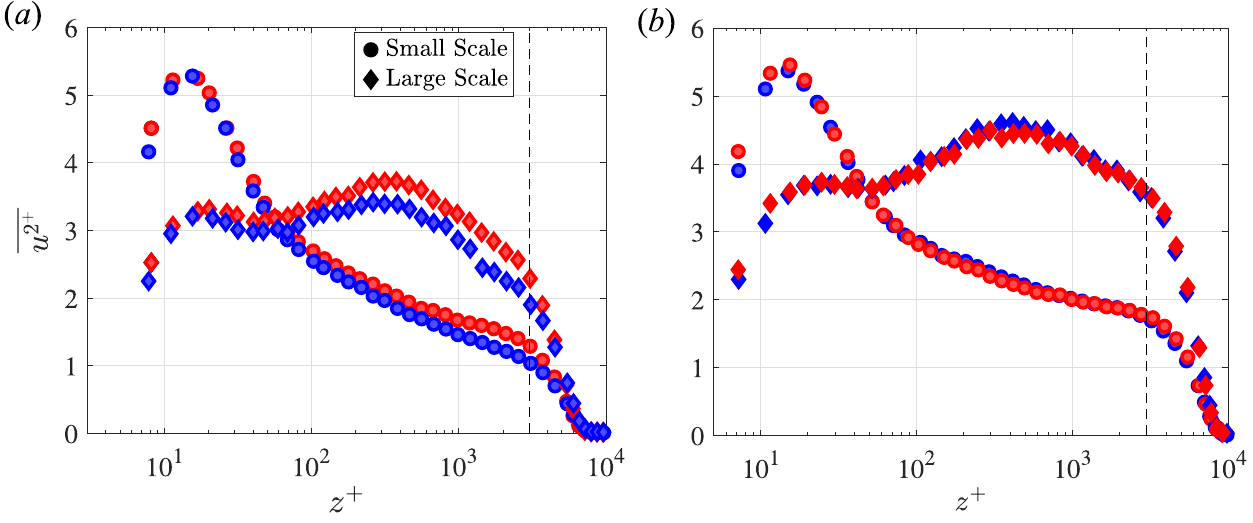}
   \end{center}
   \caption{\az{Large- and small-scale contributions to the normal stress for the high-$\ret$ Perturbed (blue), and Ref. APG3 (red) cases at (\textit{a}) $x=x_1$ and (\textit{b}) $x=x_2$, with nominally matched local flow conditions ($\ret$ and $\beta$).}}
   \label{fig8}
\end{figure}

To support these conclusions and compare them to previous studies, we decompose the fluctuating streamwise velocity component ($u$) into small-scale ($u_S$) and large-scale ($u_L$) contributions, \az{enabled by a clear separation between these scales highlighting the importance of high $\ret$ to understand such effects.} 
We compare the Perturbed and Ref. APG3 cases at $x_1$ and $x_2$, with locally matched flow conditions ($\ret \approx 8000$, $\beta \approx 0.64$; see figure~\ref{fig8}(a,b))\az{, but qualitative differences in their upstream PG histories (because of the imposed PG perturbation)}. 
The influence of $\ret$ and $\beta$ on $\overline{u_S^{+2}}$ and $\overline{u_L^{+2}}$ has been well documented \citep[e.g.][]{lozier2024revisiting, deshpande}; however, the present methodology allows us to isolate the additional contribution of PG history, independent of these parameters. 
To perform the decomposition, a cutoff timescale ($T_c^+ = 350$) was selected based on the premultiplied energy spectra, following established practices \citep{lozier2024revisiting, deshpande}, such that $u_S$ includes fluctuations with $T^+ < T_c^+$ and $u_L$ includes those with $T^+ \geq T_c^+$. 
First, the results show good collapse between the small- and large-scale contributions for both cases near the wall, particularly around the inner peak ($\zp \approx 15$). 
There are also differences in large-scale contributions between the Perturbed and Ref. APG 3 cases, across the overlap and wake regions, while differences in small-scale contributions are limited only to the wake region,\az{ which agrees with the observations of \citet{sanmiguel_vila_experimental_2020}.}  
In contrast, \az{previous APG studies at low-$\ret$ \citep{harun} have shown} large-scale differences not only in the wake region but also closer to the wall ($\zp \leq 10$), highlighting limitations originating from limited separation between these regions/scales at low-$\ret$. 
\az{These results underscore the need to continue developing frameworks that can capture these scale- and region-dependent PG history effects, which are revealed for high-$\ret$ TBLs.}

\begin{figure}
  \captionsetup{width=1.00\linewidth}
   \begin{center}
    \includegraphics[width=1.00\textwidth]{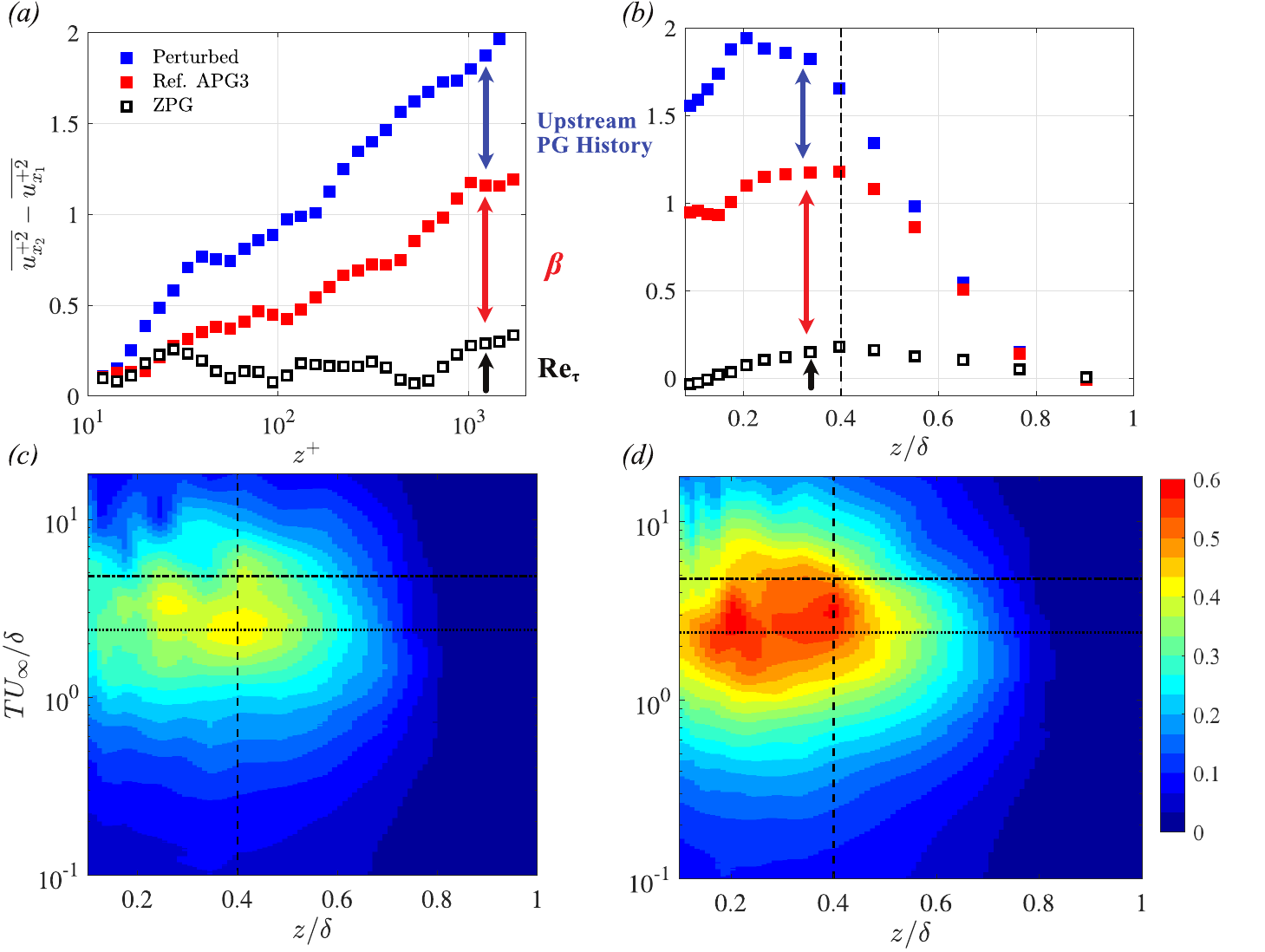}
   \end{center}
   \caption{Profiles of the change in streamwise normal stress between $x_2$ and $x_1$ for (\textit{a}) inner scaled and (\textit{b}) outer scaled wall-normal distances. Differences between the premultiplied energy spectra at $x_2$ and $ x_1$ ($f\phi_{uu,x_2}^+ - f\phi_{uu,x_1}^+$) for (\textit{c}) the Ref. APG3 case and (\textit{d}) the Perturbed case. Vertical dashed lines indicate $z/\delta=0.4$. Horizontal dot-dashed lines denote $TU_{\infty}/\delta=4.8$ and horizontal dotted lines denote $TU_{\infty}/\delta=2.4$.}
   \label{fig9}
\end{figure}

Figures~\ref{fig5} and \ref{fig7} also reveal a unique transitional behaviour in the Perturbed case between $x_1$ and $x_2$, where the turbulence statistics and energy distributions initially differ from the Ref. APG3 case, but eventually collapse after both cases experience sustained development under matched PG conditions (figure~\ref{fig3}). 
This permits an investigation of the streamwise evolution between these locations to isolate the effects of $\ret$, $\beta$, and PG history.
For each case (Perturbed, ZPG, Ref. APG3), we compute differences in the streamwise normal stress and premultiplied spectra between $x_1$ and $x_2$: $\overline{u_{i,x_2}^{+2}} - \overline{u_{i,x_1}^{+2}}$ and $\Delta_i = (f\phi_{uu}^+)_{i,x_2} - (f\phi{uu}^+)_{i,x_1}$. 
These results are shown in figure~\ref{fig9}, where black squares represent the effect of increasing $\ret$ (ZPG), red squares show the combined effects of $\ret$ and $\beta$ (Ref. APG3), blue squares capture the influence of $\ret$, $\beta$, and PG history (Perturbed), and the differences between these curves decouple each contribution. 
The normal stresses are shown in both viscous ($\zp$) and outer scaling ($z/\delta$) to highlight PG effects near the wall and in the wake region, accounting for the change in $\ret$ between $x_1$ and $x_2$.
As expected, increasing $\ret$ only raises the normal stress in the overlap and wake regions \citep[figures~\ref{fig9}a,b][]{marusic_evolution_2015}. 
The addition of APG conditions energises turbulent motions around $TU_{\infty}/\delta \approx 2.4$ and $z/\delta = 0.4$ (figure~\ref{fig9}c), which also corresponds to an increase in normal stress for the Ref. APG3 case. 
In contrast, the Perturbed case (figure~\ref{fig9}d) shows significantly greater energy growth between $x_1$ and $x_2$, with a broader distribution and the peak shifted closer to the wall ($z/\delta \approx 0.25$), suggesting a `lag' in the development/energization of APG-related scales due to its upstream PG history. 
These results also suggest a lag between APG-related effects originating primarily in the wake region and those observed in the overlap region. 
Overall, the results reinforce the need for sufficiently long APG development regions, especially under rapidly changing upstream PG conditions, to allow PG history effects to decay when studying the streamwise evolution of high-$\ret$ APG TBLs or when developing models that reliably capture local APG effects. 

\section{High-Reynolds-number APG TBLs with minimal PG history}
\label{MinHist}

\begin{figure}
  \captionsetup{width=1.00\linewidth}
   \begin{center}
    \includegraphics[width=01.00\textwidth]{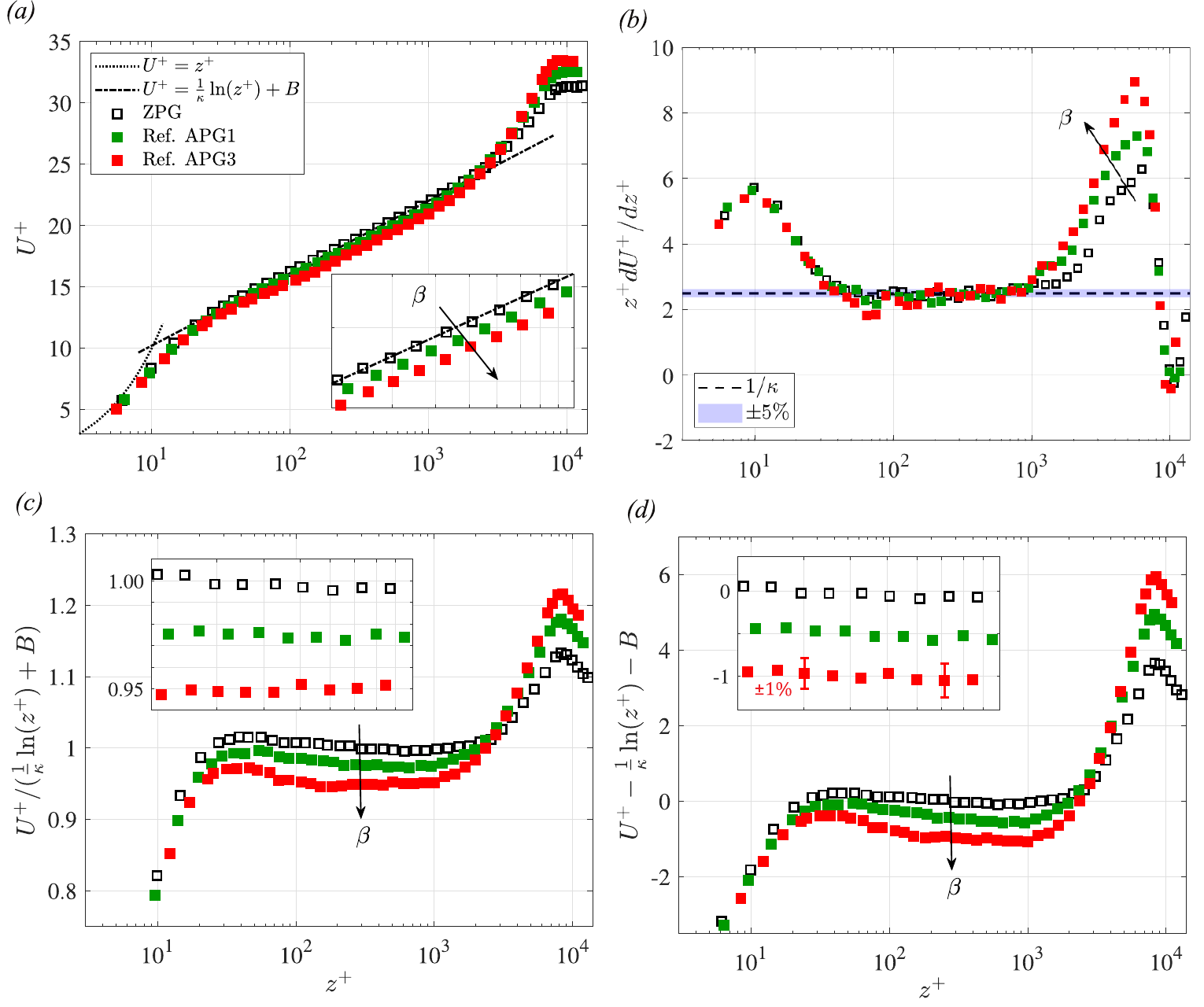}
   \end{center}
   \caption{(\textit{a}) Profiles of mean streamwise velocity for ZPG and APG cases with minimal PG histories at $x=17.4$ m. All cases have nominally matched $\ret\approx 10000$, but varying $\beta$ (see table~\ref{tab:1}). Corresponding (\textit{b}) indicator functions, (\textit{c}) ratios with the classical log-law \eqref{eq:loglaw} and (\textit{d}) deviations from the classical log-law.}
   \label{fig11}
\end{figure}

We now investigate the development of high-$\ret$ moderate-APG TBLs with minimal PG histories, i.e. strictly ZPG conditions upstream of the APG development region (figure~\ref{fig3}b). 
These experiments enable a rigorous evaluation of local APG effects on classical TBL scaling behaviours, with a sufficient overlap region and the absence of upstream PG perturbations. 
To this end, mean streamwise velocity profiles with nominally matched $\ret$ but varying $\beta$ are presented in figure~\ref{fig11}(a). 
Additionally, the indicator function, deviations from the log-law ($\up{-}\up_{log}$), and ratios with respect to the log-law ($\up/\up_{log}$ following \citealt{nagib2024method}) are shown in figures~\ref{fig11}(b–d), respectively. 
These measurements correspond to the ZPG, APG1, and APG3 cases at $x = 17.4,\rm(m)$ ($x_2$), with characteristics listed in table~\ref{tab:1}. 
Since all three cases share an identical PG history upstream of $x = 8.5$ m (figure~\ref{fig3}b), and have nominally matched local $\ret$, any differences in the mean velocity profiles or their scaling are attributable solely to the local APG. 

A key observation is the systematic deviation of the APG mean velocity profiles from the ZPG profile in the overlap and wake regions, with the magnitude of deviation increasing with $\beta$, as shown in figure~\ref{fig11}(a). 
Similar qualitative trends have been reported in both experimental \citep{nagano1998structure, deshpande} and numerical studies \citep{lee_structures_2009, yoon, bobke2017history, pozuelo, gungor2024turbulent}; however, only the present high-$\ret$ dataset, with well-controlled PG histories and clear scale separation, enables this deviation to be quantified precisely and unambiguously, including its sensitivity to PG history. 
Figure~\ref{fig11}(c) shows that the ZPG mean velocity profile agrees with the classical logarithmic scaling law to within 1\% across the entire overlap region, while the APG3 case (highest $\beta$) deviates by more than 5\%, a statistically significant difference given the experimental uncertainty. 
These deviations are also systematic, as the intermediate APG1 profile falls between the ZPG and APG3 profiles. 
However, such deviations in the ratio profiles can be attributed to changes in either $\kappa$ and/or $B$. 
To investigate these coefficients independently, $\kappa$ is first isolated using the indicator function profiles in figure~\ref{fig11}(b). 
A broad wall-normal region with nominally constant indicator function magnitude confirms robust logarithmic scaling behaviour for all cases, and no systematic or statistically significant change in $\kappa$ is observed with increasing $\beta$ (for the high-$\ret$ low-to-moderate APG cases considered). 
The additive coefficient $B$ is then isolated using the subtraction profiles in figure~\ref{fig11}(d), which show a significant and systematic decrease in the magnitude of $B$ with increasing $\beta$. 

\begin{figure}
  \captionsetup{width=1.00\linewidth}
   \begin{center}
    \includegraphics[width=0.85\textwidth]{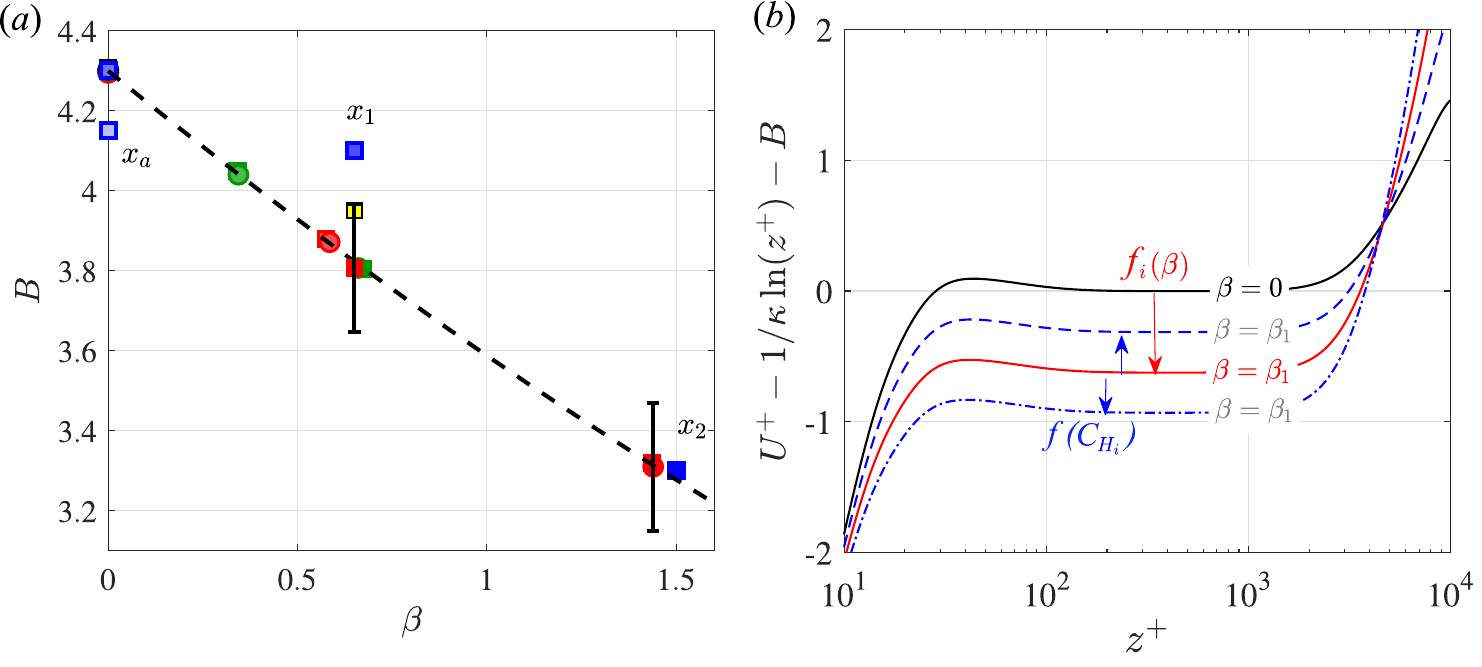}
   \end{center}
   \caption{\az{(\textit{a}) Variation of the classical log-law \eqref{eq:loglaw} additive coefficient, $B$, as a function of $\beta$ for high-$\ret$ low-to-moderate APG TBLs with minimal PG history effects (see figure~\ref{fig3}b and table~\ref{tab:1}). Note that the circle and square symbols with the same colour correspond to matched $\beta$ but different $Re_{\tau}$. The yellow squares represent the Weakly Perturbed case, see appendix~\ref{apx}. (\textit{b}) A schematic of mean velocity shift from the classical log-law for APG cases with matched $\beta$ but different PG histories.}}
   \label{fig12}
\end{figure}

Due to the systematic decrease in $B$ observed for the highest-$\beta$ cases at $x = 17.4$ m, the additive coefficient was estimated for all present cases (see table~\ref{tab:1}). 
Deviations from the classical value of $B = 4.3$ were quantified from the subtraction profiles (figure~\ref{fig11}d) by averaging over the wall-normal region where the profiles remain approximately constant (as shown in figure~\ref{fig11}b). 
The resulting values of $B$ as a function of $\beta$ are presented in figure~\ref{fig12}. 
These results confirm a systematic decrease in $B$ with increasing $\beta$, while also indicating that $B$ remains largely independent of $\ret$ across the range considered.

According to the high-$\ret$ framework proposed by \citet{marusic_logarithmic_2013}, both $\kappa$ and $B$ are considered universal or Reynolds number invariant for the three canonical flows (ZPG TBL, pipe, and channel); however, this `universality' is an active topic of research \citep{nagib_variations_2008, mon23}. 
The new experimental results presented here, obtained under systematically varying PG at high $\ret$, indicate that such universality does not necessarily extend to non-canonical flows. 
Our results demonstrate that while the value of $\kappa$ remains invariant \citep[within experimental uncertainty and for the low-to-moderate APGs, consistent with the findings of ][]{knopp2021experimental}, the additive constant $B$ varies with the imposed boundary conditions. 
Specifically, since the TBL boundary conditions are altered by streamwise pressure gradients, it is expected that $B$ should vary systematically with both $\beta$ and PG history. 
This observation aligns with previous experimental \citep{nagano1998structure, monty2011} and numerical \citep{bobke2017history, gungor2024turbulent} studies reporting reductions in $B$ under APG conditions. 
However, many earlier investigations were conducted at relatively low Reynolds numbers and/or under significant PG history effects, and experimental studies have not always determined $U_{\tau}$ via independent measurements, limiting the ability to draw conclusive trends. 

\az{
\section{Towards characterizing PG history effects}
\label{sec:comparisons}

Here, we contextualize the present dataset and results, featuring high~$\ret$ TBLs with moderate~$\beta$ and well-controlled PG histories, towards the development of new composite mean velocity profile formulations for APG TBLs. 
The indicator function ($z^+ \mathrm{d}U^+/\mathrm{d}z^+$) for the dataset of \citet{pozuelo}, taken at similar $\beta$ but lower~$\ret$ relative to the present study, and for the dataset of \citet{knopp2021experimental}, taken at similar $\ret$ but stronger~$\beta$, are shown in figure~\ref{fig13}(a). 
In addition, the dataset of \citet{romero_properties_2022}, which has nominally matched $\ret$ and $\beta$ with the present cases at $x_1$ (but with unique PG history effects), is included in figure~\ref{fig13}(c).

From figure~\ref{fig13}(a), it is difficult to confirm the existence of a region with logarithmic scaling for the low-$\ret$ case owing to insufficient separation between the inner and wake regions. 
Similarly, at high~$\ret$, increasing $\beta$ significantly amplifies the wake region, causing the indicator function to depart from the plateau ($1/\kappa$) earlier than in the moderate-$\beta$ case. 
Finally, figure~\ref{fig13}(c) shows that, even under nominally matched high-$\ret$ and moderate-$\beta$ conditions, varying PG history effects can make it difficult to determine the existence and/or extent of the overlap region and the values of the log-law coefficients.

\begin{figure}
  \captionsetup{width=1.00\linewidth}
   \begin{center}
    \includegraphics[width=1.0\textwidth]{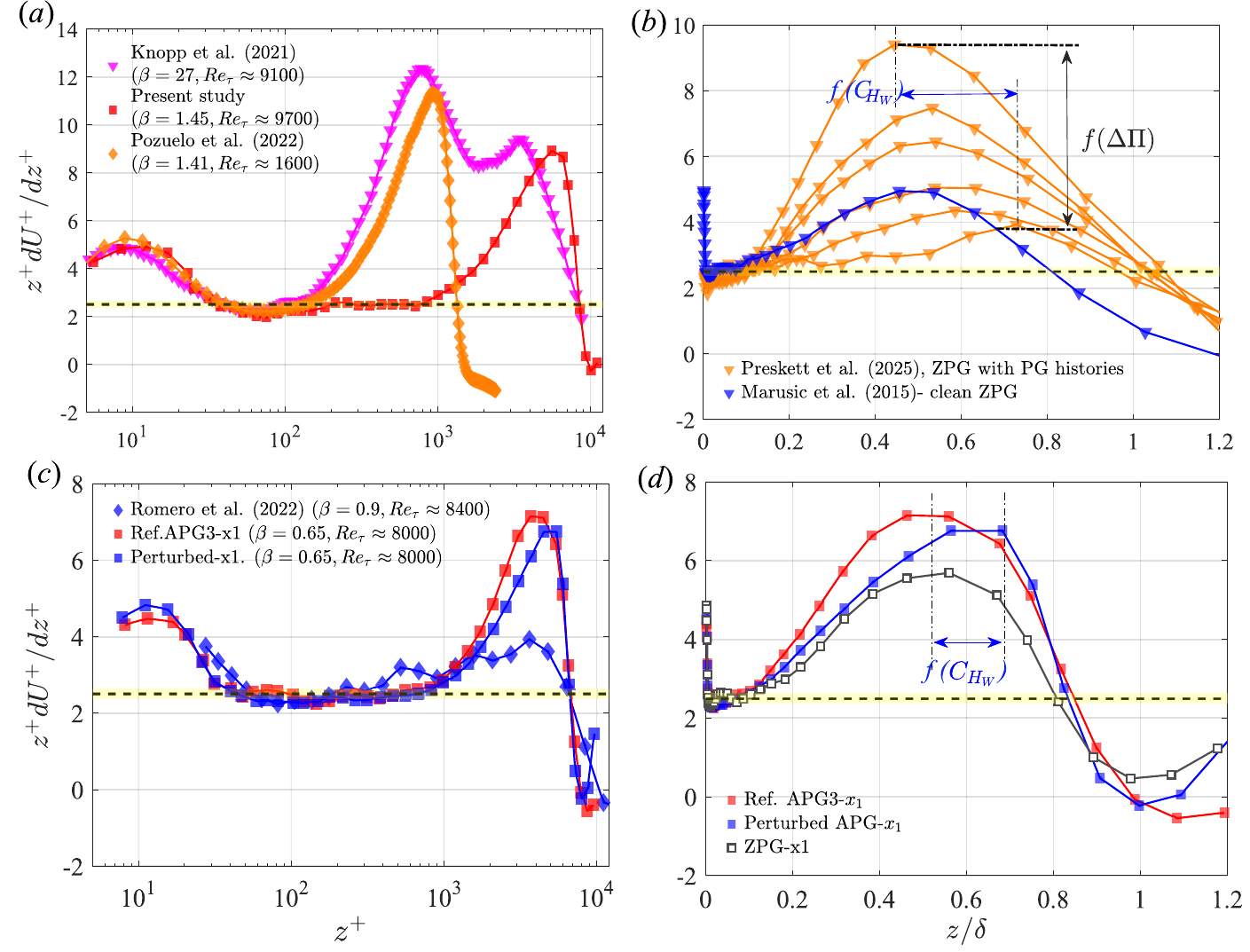}
   \end{center}
   \caption{\az{Indicator functions of selected cases from (\textit{a}, \textit{c}, \textit{d}) the present dataset at $x=x_1$, the low-$\ret$, matched-$\beta$ dataset of \citet{pozuelo}, the matched high-$\ret$ high-$\beta$ dataset of \citet{knopp2021experimental}, and the matched $\ret$ and $\beta$ dataset with PG history effects of \citet{romero_properties_2022}; (\textit{b}, \textit{d}) the high-$\ret$ ZPG dataset with varying PG history of \citet{preskett2025effects}, and the reference ZPG dataset of \citet{marusic_evolution_2015}. The horizontal dashed line indicates $1/\kappa$, with the $\pm 5\%$ range shown by the shaded yellow region.}}
   \label{fig13}
\end{figure}

\subsection{PG history effects in the wake region}

It can be argued that the wake region is the part of the TBL most directly affected by both local PGs and PG history; therefore, a clear examination of the wake region is essential for identifying/quantifying PG history effects. 
To demonstrate this, we calculate the indicator function for the datasets of \citet{preskett2025effects}, specifically for TBLs with locally ZPG conditions and matched inlet conditions, but with distinct PG histories (figure~\ref{fig13}b). 
The corresponding indicator functions for the present cases at $x_1$ are also shown in figure~\ref{fig13}(d). 
In both datasets, there are differences in the magnitude \emph{and} shape of the wake region profiles that are solely attributable to differences in PG histories, since all other flow conditions are matched. 
First, different PG histories lead to significant differences in the wake strength, $\Pi$, and \citet{preskett2025effects} also recognized and quantified this effect using the deviation of $\Pi$ from its ZPG value, $\Delta\Pi = \Pi - \Pi_{\rm ZPG}$. 
However, the wake region profile is also `stretched' under the influence of different PG histories, as indicated by the shift in the peak of the indicator function relative to the reference ZPG dataset at matched $\ret$ \citep{marusic_evolution_2015}. 
This suggests that characterizing PG-history effects requires accounting for more than changes in $\Pi$ alone, and that an additional parameter is needed to capture this wall-normal stretching. 
This is highlighted in figure~\ref{fig13}(b,d) by a \textit{history-effect parameter} for the wake region ($C_{H_W}$), where $H_W$ denotes the history contribution in the wake region. 
Quantification and further discussion of this parameter will be presented in Part~2.

\subsection{PG history effects in the overlap region}

As discussed earlier, at both low and high $\ret$, the mean velocity in the overlap region exhibits a measurable shift from the classical log-law when a local APG is imposed. 
The PG history further influences this shift \citep{bobke2017history, pozuelo}, even when the local flow conditions are nominally ZPG \citep{gungor2024turbulent}. 
The direction and magnitude of this shift vary across datasets, leading to either a relatively larger or smaller displacement than that observed for the present APG cases with minimum PG history, figure~\ref{fig12}(a). 
A schematic illustration of the deviation of the mean velocity from the classical log-law is provided in figure~\ref{fig12}(b), highlighting the behaviour of cases with minimal PG history effects (where $B$ is a function of $\beta$ only) and for cases with the additional influence of PG history. 
This suggests that an additional parameter is needed to account for the influence of PG history on this deviation, a \emph{history effect factor} for the inner region, $C_{H_{\mathrm{i}}}$, where $H_i$ denotes the history contribution in the inner region. 
These combined observations of the behaviours of $B$ and the wake region profile, due to local and upstream PGs, for a range of datasets in the literature provide a foundation for developing a new composite-fit formulation which accounts for the local conditions ($\ret$ and $\beta$) while also capturing the influence of the history effect, which will be presented in Part 2 \citep{zarei2026_part2}.}

\section{Conclusions}

\az{This study experimentally decouples and quantifies the influence of upstream pressure-gradient (PG) history on high-Reynolds-number turbulent boundary layers subjected to low-to-moderate adverse pressure gradients (APGs), using well-resolved hot-wire and OFI measurements. 
The primary aim was to address persistent uncertainties in the literature by rigorously assessing the universality of the von Kármán ($\kappa$) and additive ($B$) coefficients in the classical logarithmic scaling law for the mean velocity profile of high-$\ret$ APG TBLs. 
Leveraging a recently developed methodology that generates APG TBLs with prescribed upstream PG histories, we find that the von Kármán coefficient $\kappa$ remains invariant within experimental uncertainty, whereas the additive constant $B$ varies systematically with both the local APG strength and the upstream PG history. 
These observations provide a possible explanation for the variability in logarithmic scaling behaviours reported in previous studies, which may have been influenced by a combination of significant PG history effects, low Reynolds numbers and/or uncertainties in the friction velocity.

The experimental facility and methodology employed here enabled precise control of both the location and magnitude of perturbations in the upstream region while maintaining well-matched conditions with the reference APG TBL (with minimal PG history) in the downstream section. 
The present study focused on a deliberately weak upstream APG perturbation followed by an extended relaxation region with nominally ZPG conditions, leading to a downstream development region characterised by sustained, matched high-$\ret$ and moderate APG conditions. 
This approach enabled the systematic decoupling of PG history effects from the local flow parameters ($\ret$ and $\beta$) and allowed their individual influences on turbulence statistics and premultiplied energy spectra to be examined across different regions of the boundary layer.

Importantly, detectable differences emerge even under these weak variations in PG history, indicating that the same effects are likely to be present—and potentially more pronounced—under stronger perturbations or shorter relaxation distances. 
While the present study does not directly extrapolate its results to stronger APGs or more complex upstream PG sequences, the controlled high-$\ret$ environment and long development region ensure that the observed sensitivity to PG history represents a robust physical behaviour relevant to a broader range of APG TBL configurations. 
The high-$\ret$ conditions also provide sufficient scale separation to rigorously investigate classical logarithmic scaling and to examine how turbulent motions across different scales and regions of the TBL respond to PG history.

In the relaxation region, the results confirm that upstream PG perturbations continue to influence the flow even after the local PG ($\beta$) returns to zero, corroborating earlier low-$\ret$ findings \citep{bobke2017history, gungor2024turbulent}. 
However, after sufficiently long relaxation distances (e.g.\ $26\delta$), the mean velocity and normal stress profiles recover classical ZPG behaviour, while other characteristics such as boundary-layer thickness and growth rate remain elevated relative to the reference ZPG case. 
In the downstream development region, even under nominally identical local conditions (matched $Re_\tau$ and $\beta$), differences in immediate PG history upstream of the measurement location significantly affect the mean velocity profile, particularly within the logarithmic region where classical scaling laws apply.

The measurements further show that local APGs influence both small- and large-scale turbulent motions in the wake region ($z \approx 0.4\delta$), whereas PG history primarily affects large-scale motions and extends farther toward the overlap region ($z \approx 0.25\delta$). 
At high $\ret$, the increased scale separation indicates that PG-related effects remain largely confined to the overlap and wake regions, while near the wall small-scale motions remain essentially unaffected, in contrast to previous observations at lower $\ret$. 
These results suggest a scale-dependent adjustment to changing PG conditions, with large-scale motions responding more slowly and the adjustment propagating from the wake region toward the overlap region.

Cumulatively, these findings highlight the importance of establishing experimental TBLs with well-controlled upstream PG histories when systematically investigating the classical log-law behaviour of high-$\ret$ APG TBLs. 
The results further indicate that a single parameter is insufficient to quantify or model PG-history effects on the mean velocity profile. 
In the wake region, changes in $\Pi$ quantify the vertical displacement of the mean profile due to PG history \citep{preskett2025effects}, but do not capture changes in the wake profile shape or horizontal stretching. 
Similarly, in the overlap region, PG history introduces additional deviations from the $\beta$-based trend in the log-law additive constant $B$. 
These observations therefore suggest that at least two independent parameters—one governing the inner/overlap regions and another governing the wake region—are required for a physically consistent description of the mean velocity profile in APG TBLs with varying PG histories.

The measurements presented here provide a high-quality dataset and new physical insight into the roles of $\ret$, local PGs and PG history in shaping the mean velocity profile. 
In particular, the results demonstrate that while the von Kármán coefficient $\kappa$ remains universal at high Reynolds numbers, the additive constant $B$ depends on both the local APG strength and the upstream PG history. 
Together with existing datasets in the literature, and drawing inspiration from the framework of \citet{nickels}, these results establish the physical basis for developing an improved composite description of the mean velocity profile for APG TBLs. 
Building on the mechanisms identified here, Part~2 introduces and tests a composite mean velocity profile formulation that explicitly incorporates the effects of both local pressure gradients and PG history.}

\backsection[Acknowledgments]
{\noindent The authors gratefully acknowledge funding from the Office of Naval Research (ONR) and ONR Global; Grant No. N62909-23-1-2068.
R. Deshpande also acknowledges financial support from the Melbourne Postdoctoral Fellowship awarded by the University of Melbourne, and is grateful to A/Prof R. Vinuesa for engaging discussions on APG TBLs.}

\backsection[Declaration of Interests]{The authors report no conflict of interest.}

\appendix
\section{Weakly Perturbed case}
\label{apx}
\begin{figure}
  \captionsetup{width=1.00\linewidth}
   \begin{center}
    \includegraphics[width=0.75\textwidth]{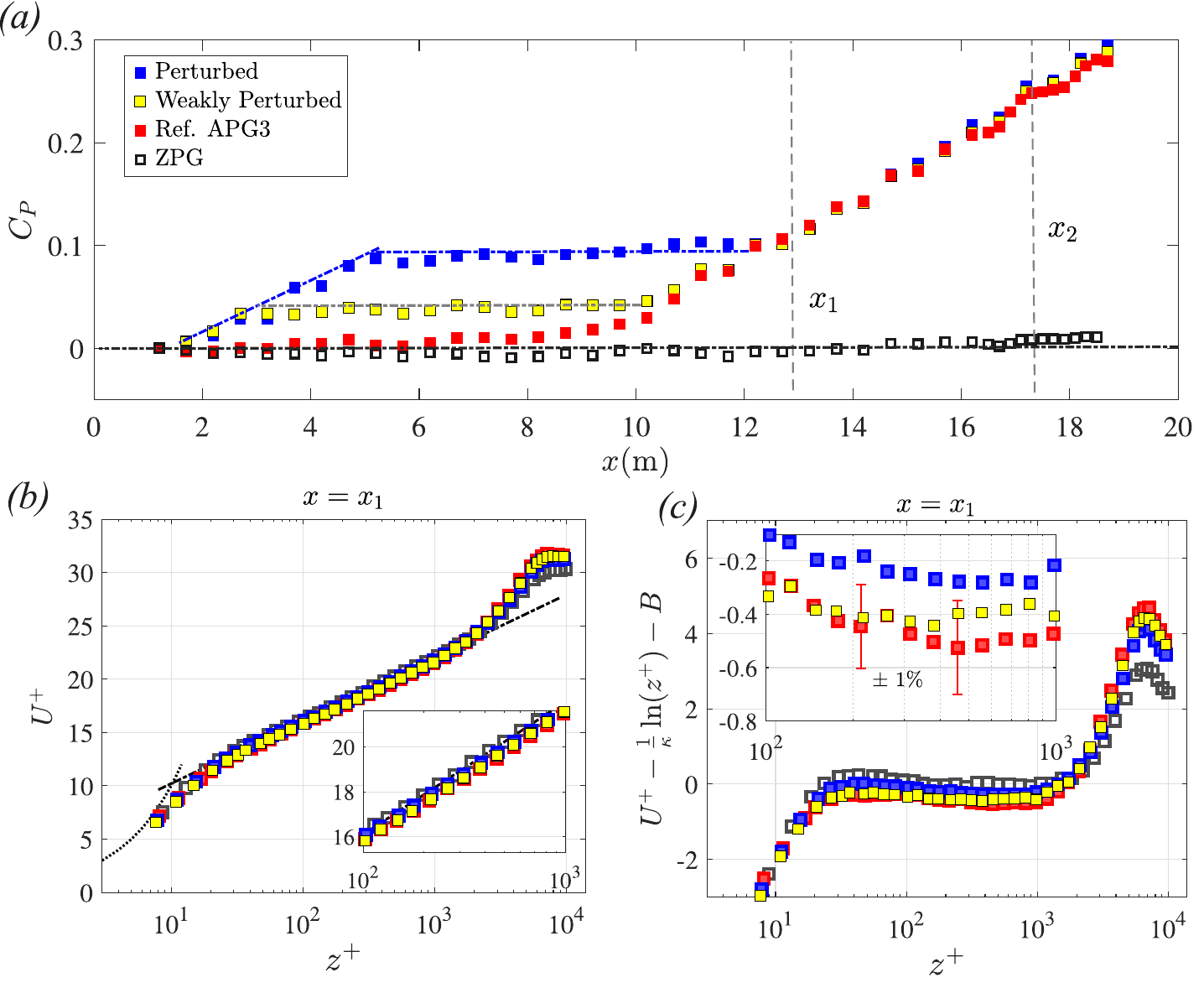}
   \end{center}
   \caption{(\textit{a}) $C_P$ profiles for all cases, including Perturbed and Weakly Perturbed cases with distinct upstream APG histories, (\textit{b}) mean streamwise velocity profiles and (\textit{c}) deviations from the classical log-law.}
   \label{figA1}
\end{figure}

Figure~\ref{fig3} presents a controlled APG perturbation introduced in the upstream region to strategically modify the pressure-gradient history of the Perturbed case, in comparison to the reference APG3 case. 
Additionally, we examined a second case with a relatively weaker upstream APG perturbation (the “Weakly Perturbed” case) to investigate the minimum perturbation strength necessary to observe PG history effects in the turbulence statistics at $x = x_1$. 
The $C_P$ profiles of these two Perturbed cases are shown in figure~\ref{figA1}(a), along with the Ref. APG3 and ZPG cases. 
At $x_1$, $\ret$ and $\beta$ are still nominally matched across all cases, including the Weakly Perturbed case.
The mean velocity profiles and their deviations from the classical logarithmic scaling law are presented in figures~\ref{figA1}(b,c). 
The mean velocity and deviation profiles (in the overlap region) for the Weakly Perturbed case fall between those of the Ref. APG3 and Perturbed cases, showing a systematic trend consistent with the earlier discussion. 
However, the differences between the Weakly Perturbed and Ref. APG3 cases fall within experimental uncertainty. 
As such, the Weakly Perturbed case was not considered in the main analysis.

\az{
\section{The configuration of the opened and closed slats}
\label{apxC}
In figure~\ref{figA2}, the configuration of the opened and closed slats used to generate the Perturbed case is shown, together with the locations of the critical measurement stations $x_a$ and $x_1$. 
As illustrated, $x_a$ lies between closed slats, multiple $\delta$ downstream of the transition to ZPG conditions. 
Similarly, $x_1$ lies between open slats, multiple $\delta$ downstream of the transition to APG conditions. 
Although this PG perturbation is introduced as a step change, the $x_a$ and $x_1$ locations lie several boundary-layer thicknesses downstream of this step change (i.e. beyond the classical step-response length of $O(\delta)$), where the local flow conditions have re-stabilised, confirmed by the plots of $C_P$ in figure~\ref{fig3}(a). 
In this way, the turbulence statistics at $x_a$ and $x_1$ capture the persistence of the influence of the upstream PG conditions after the local conditions have changed. 
Further, the freestream at $x_a$ matches the freestream velocity at $x_b$ and $x_c$, validating that this station is indeed located within the ZPG region. 
Similarly, the freestream velocity for both the Perturbed and Ref. APG3 cases at $x_1$ are matched, indicating that this station is within the APG region for the Perturbed cases. 
Additional evidence supporting that $x_1$ has locally APG conditions comes from the fact that the flow has ZPG-like behaviour at $x_c$, while the mean velocity and turbulence statistics at $x_1$ deviate from these ZPG trends (exhibiting a strengthened wake, a shift in the overlap region, and differences in second-order statistics). 

\begin{figure}
  \captionsetup{width=1.00\linewidth}
   \begin{center}
    \includegraphics[width=0.75\textwidth]{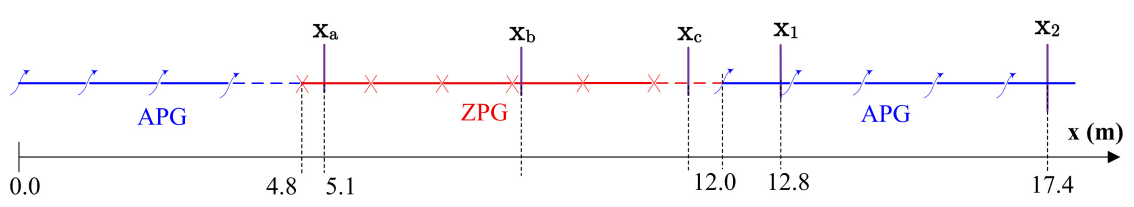}
   \end{center}
   \caption{Exact configuration of slat openings/closures and the corresponding measurement stations for the Perturbed and reference cases.}
   \label{figA2}
\end{figure}
}

\bibliographystyle{jfm}
\bibliography{references}

\end{document}